\documentclass[
elsearticle,
 amsmath,amssymb,
 aps,
]{revtex4-2}

\usepackage{graphicx}
\usepackage{dcolumn}
\usepackage{bm}

\input{myPreamble.tex}

\begin{document}

\preprint{APS/123-QED}

\title{Turbulent puffs in transitional pulsatile pipe flow at moderate pulsation amplitudes}

\author{Daniel Mor\'on}
\affiliation{%
 University of Bremen, Center of Applied Space Technology and Microgravity (ZARM),
Am Fallturm 2, 28359 Bremen, Germany.
}%
\author{Marc Avila}%
\affiliation{%
  University of Bremen, Center of Applied Space Technology and Microgravity (ZARM),
Am Fallturm 2, 28359 Bremen, Germany.
}%
 \affiliation{University of Bremen, MAPEX Center for Materials and Processes,
Am Biologischen Garten 2, 28359 Bremen, Germany.}
\date{\today}

\DeclareRobustCommand{\mod}[1]{{#1}\xspace}
\DeclareRobustCommand{\modd}[1]{{#1}\xspace}

\begin{abstract}
We show that, in the transitional regime of pulsatile pipe flow, at moderate-to-high amplitudes $0.5 \lesssim \Amplitude \lesssim 1$, the first long-lived turbulent structures are localized and take the form of the puffs and slugs observed in statistically steady pipe flow. We perform direct numerical simulations at many pulsation frequencies, amplitudes and \Reynolds, and observe different dynamics of puffs and slugs. At certain flow parameters we find, using a causal analysis, that puffs actively make use of linear instabilities in the laminar \mod{Sexl-Womersley} profile to survive the pulsation. Using all these lessons learned, we extend a low order model by Barkley \textit{et al.}, Nature (2015), to reproduce these dynamics. We find a good agreement between the extended model and our numerical results in a broad parametric space of pulsation amplitudes $0.5 \lesssim \Amplitude \lesssim 1$, frequencies $\Womersley \gtrsim 5$ and $2100 \leq \Reynolds \leq 3000$. With the help of our numerical results, causal analysis and model, we determine that turbulence production has two sources at these flow parameters: the mean shear as in statistically steady pipe flow, and the instabilities of the instantaneous pulsatile mean profile.
\end{abstract}

\maketitle

\section{Introduction}
More than a century after Reynolds experiments \citep{Reynolds}, we are closer to fully understand turbulence transition in statistically steady pipe flow (SSPF), see \citet{AvilaAnnRev} for a recent review. For sufficiently perturbed flows, transition only depends on the Reynolds number $\Reynolds=\sfrac{UD}{\nu}$, where $U$ is the time averaged bulk velocity, $D$ the diameter of the pipe and $\nu$ the kinematic viscosity of the fluid. Despite being linearly stable, at least up to $\Reynolds \leq 10^{7}$ \citep{Meseguer}, perturbations in SSPF can make use of non-modal mechanisms to grow, saturate and trigger turbulence at much lower $\Reynolds \sim \mathcal{O} \left( 10^{3} \right)$ \citep{eckhardt}. The optimal perturbation, the one that attains a higher energy growth in SSPF, is a pair of streamwise vortices \citep{schmid2000stability}. 

Once triggered, turbulence in SSPF first appears in the form of localized turbulent patches of constant length, known as turbulent puffs. Depending on the Reynolds number, puffs are more likely to either decay (at low $\Reynolds \lesssim 2040$), split (at $2040 \lesssim \Reynolds \lesssim 2250$) or elongate into slugs (at $\Reynolds \gtrsim 2250$) \citep{AvilaAnnRev}. Also depending on \Reynolds, puffs (slugs) move (and elongate) at a certain upstream (and downstream) front speed $c_{u}$ $\left(\text{and }c_{d}\right)$. The exact mechanisms by which turbulent puffs decay, split or elongate are still un-clear.

The low-order model proposed by \citet{barkley2015rise} (hereafter referred to as the Barkley Model, BM) is able to reproduce the front speeds of turbulent puffs and slugs in SSPF. In this paper we provide a description of the model in appendix~\ref{ap:ebm}. The reader is referred to \citet{barkley2016} and references therein for a more detailed description. In short, the model considers only two one-dimensional, time-dependent variables, $q\left(x,t\right)$ and $u\left(x,t\right)$, whose evolution is described by two non-linearly coupled advection-diffusion-reaction equations. These equations are inspired by, but not derived from, the Navier--Stokes equations. In fact, the variables and the parameters represent features of pipe flow, but have arbitrary physical units. The variable $q$ represents the turbulence intensity, and $u$ the state of the local mean shear of the flow, at each axial location $x$ and time $t$. The key feature of the BM is the non-linear interaction between $u$ and $q$. The turbulence intensity $q$ takes advantage of the mean shear $u$ to grow. However, in the axial locations where $q>0$, the local mean shear is reduced \cite{kuhnen}, and in turn, adversely affects the growth of $q$. When fitted correctly, the model returns the turbulent front speeds $c_{u}$ and $c_{d}$ of turbulent structures in SSPF with high accuracy in a broad \Reynolds regime \citep{barkley2015rise,Song2017,chen2022}. The remarkable success of this model has motivated some researchers to use it to study puff split dynamics, \citep{splitBarkley}, or even turbulence transition of non-Newtonian pipe flow \citep{heatedPipe}. The question still remains of, to what extent, the assumptions and simplifications of the BM are correct, and if it can be easily adapted to similar flow set ups, such us pipe flows driven at an unsteady flow rate. 
We study pulsatile pipe flows, whose bulk velocity $\bar{U}\left(t\right)$ has one harmonic component: 
\begin{equation}
    \bar{U}\left(t\right)=U\left[1+A\sin\left(2 \pi f \cdot t\right) \right] \text{.}
    \label{eq:pulsse}
\end{equation}
The flow depends on three parameters: the mean \Reynolds, the frequency of the pulsation $f=1/T$ or its non dimensional counterpart, the Womersley number $\Womersley=\sfrac{D}{2}\sqrt{\sfrac{2\pi f}{\nu}}$, and the amplitude \Amplitude. In the case of smooth rigid pipes, considered here, there is an analytical solution to laminar pulsatile pipe flow, the Sexl-Womersley (SW) profile $U_{SW}\left(r,t\right)$ \mod{\cite{Sexl1930,womersley1955method}}.
At small-to-moderate amplitudes $\Amplitude \leq 0.4$ and transitional $\Reynolds \approx 2000$, the first long-lived turbulent structures found in pulsatile pipe flows are also turbulent puffs \citep{xu2017}. Their behavior depends on the pulsation frequency. At high $\Womersley\gtrsim20$ (independently of \Amplitude) puffs do not have enough time to adapt to the fast harmonic driving \citep{xu2018}, and their behavior is identical to the one found in SSPF. At low $\Womersley\lesssim 4$ the behavior is quasi-steady, and the puff dynamics depend on the instantaneous Reynolds number $\Reynolds_{i} \left(t\right)=\frac{\bar{U}\left(t\right)}{U}\cdot \Reynolds$ \citep{xu2017}. At intermediate $5\lesssim \Womersley\lesssim 19$, the behavior of puffs smoothly transitions between the two limiting cases described above. \mod{It is still unclear} whether this also happens at $\Amplitude \geq 0.5$. 
At moderate-to-large amplitudes $0.5 \leq \Amplitude \leq 1$, $\Reynolds \approx 2000$, and intermediate frequencies, $5\lesssim \Womersley\lesssim 19$, \mod{the SW profile} is instantaneously unstable at some phases of the pulsation period \citep{moron2022}. This instability is linked to the presence and behavior of inflection points in the laminar profile \citep{cowley1987high,nebauer2019}. Helical perturbations can take advantage of this instability to grow, and trigger turbulence, as first reported by \citet{xu2020PNAS}. At these flow parameters, although the optimal perturbation of SSPF, the pair of streamwise vortices, can substantially grow in energy, helical perturbations exhibit the maximum growth \citep{xu2021non,pier_schmid_2021,kern2021}, consistent with experiments \citep{xu2020PNAS}. \mod{In DNS, helical perturbations rapidly grow, saturate and trigger turbulence that, at some flow parameters, takes} the form of localized puffs modulated in length and magnitude by the pulsation \citep{moron2022}. Recent results suggest that these turbulent puffs actively make use of the instantaneous instabilities of the \mod{SW} profile to survive during the phases of the period when $\Reynolds_{i} \left( t\right)$ is too low \citep{entropy2021}. This hypothesis however, has not been rigorously verified yet. 
In this paper, we study turbulent puffs in transitional pulsatile pipe flow at $2100 \leq \Reynolds \leq 3000$, moderate-to-large amplitudes $0.5 \leq \Amplitude \leq 1$ and intermediate frequencies, $5\lesssim \Womersley\lesssim 19$. We perform direct numerical simulations (DNS) at many \Reynolds, \Womersley and \Amplitude, and study the behavior of turbulent puffs at these flow regimes. We perform causal analyses to determine if they make use of the instabilities of the laminar profile to survive the pulsation, as suggested by \mod{Feldmann \textit{et al.}} \citep{entropy2021}. Using the lessons learned from these analyses, we extend the BM to a new Extended Barkley Model (EBM) that \mod{reproduces the dynamics of} turbulent puffs in pulsatile pipe flow \mod{throughout the parameter regime studied in the DNS}. 

The paper is organized as follows. In \S \ref{sec:methods} we describe the numerical methods we use to perform DNS of pulsatile pipe flow. In \S \ref{sec:DNSresults}, we \mod{analyze} our DNS results of pulsatile pipe flows at several flow parameters. In \S \ref{sec:master-slaveresults}, we \mod{present} the results of our causal analysis and in \S \ref{sec:ebmresults} the comparison between the extended model and DNS results. Finally, in \S \ref{sec:conclusion} we draw some conclusions.

\section{Methods} \label{sec:methods}
We consider a viscous Newtonian fluid with constant properties in a straight smooth rigid pipe of circular cross-section, with a time-dependent bulk velocity, eq.~\eqref{eq:pulsse}. The flow is assumed to be incompressible and governed by the dimensionless Navier--Stokes equations (NSE):
\begin{equation}
\frac{\partial\pmb{u}}{\partial t} + \left(\pmb{u} \cdot \nabla \right) \pmb{u} =
-\nabla p + \frac{1}{\Reynolds}\nabla^{2}\pmb{u}+P_{G}\left(t\right)\cdot\pmb{e}_{x}
\quad\text{and}\quad
\nabla\cdot\pmb{u} = 0
\text{.}
\label{eq:NSeq}
\end{equation}
Here, $\pmb{u}$ is the fluid velocity, $p$ the pressure and $P_{G}$ the time-dependent axial pressure gradient which drives the flow at the bulk velocity defined in eq.~\eqref{eq:pulsse}. All variables in this study are rendered dimensionless using the pipe diameter ($D$), the time-averaged bulk velocity ($U$) and the fluid density ($\rho_{f}$).



\subsection{Computational methods}\label{sec:NumMeth}
We perform direct numerical simulations of eq.~\eqref{eq:NSeq} using our open-source pseudo-spectral simulation code \nsp, \citep{LOPEZ}. In \nsp, the governing equations are discretized in cylindrical coordinates $\left(r,\theta,x\right)$ using a Fourier--Galerkin ansatz in $\theta$, \mod{with wavenumber $m \in \mathbb{Z}$,} and $x$ \mod{with wavenumber $\alpha=\alpha_{0} k$. Here $\alpha_{0}=2\pi/L$ where $L$ is the pipe length and $k \in \mathbb{Z}$}. \mod{We use high-order finite differences in $r$ with stencils of length 7 for master-slave simulations (see \S \ref{sec:master-slaveresults}), and of length 9 in individual DNS simulations.} Periodic boundary conditions are imposed in $\theta$ and $x$, and no-slip boundary conditions in the solid pipe wall. The discretized NSE are integrated forward in time using a second-order predictor-corrector method with variable time-step size ($\Delta t$)\mod{, as in \opipe \cite{openpipeflow}}. Further details about the \mod{implementation} of \nsp are given in \citet{LOPEZ} and references therein.

We perform single DNS of pulsatile pipe flow at different \Reynolds, \Womersley and \Amplitude. See the parameters of all our simulations at the end of the manuscript, in table~\ref{tab:dns}. The time step size is always $\Delta t < 0.0025 D/U$, the length of the pipe is $L=100D$, and the simulation with the coarsest grid in terms of $+$ units has $\num{0.044}\le\Delta r^+\le\num{2.39}$, $D^+\Delta\theta/2=\num{4.08}$ and $\Delta x^+=\num{8.27}$. We initialize the simulations with the corresponding SW profile in the whole domain. We trigger a single turbulent puff in each simulation by introducing the corresponding optimal perturbation localized in a $5D$ axial section of the pipe and scaled to $\left|\mathbf{u}^\prime_{0}\right|\approx\num{3e-2}$, following \citet{entropy2021} and \citet{moron2022}. We compute the optimal perturbation using a transient growth analysis (TGA). The reader is referred to \citet{xu2021non,moron2022} for more details on the transient growth analysis.

\mod{\subsection{Averages}\label{sec:notation}
We compute averages of the three velocity components $u_{r}$, $u_{\theta}$ and $u_{x}$ and the axial (streamwise) vorticity $\omega_{x}$. Angled brackets $\lla \bullet \rra_{\psi}$ denote averaging with respect to $\psi$, where $\psi$ stands for one or more variables. Spatial averages are performed with respect to one or more coordinates: radial $\psi \equiv r$, azimuthal $\psi\equiv \theta$, and axial $\psi \equiv x$. Temporal averages are performed using the whole time series, and are denoted with $\psi \equiv t$. Ensemble averages, using $N_{i}$ individual simulations, are denoted with $\psi \equiv N_{i}$. Phase averages correspond to averaging at several phases of the period and are denoted as $\psi \equiv t^{*}$. In the latter case, we split the pulsation period in $200$ equi-spaced period phases and we perform averages at each of them. The resulting signal is phase $\left( t^{*}\right)$ but not time $\left( t\right)$ dependent.}

\begin{figure}
\centering
\includegraphics[width=\textwidth, trim=0mm 0mm 0mm 0mm, clip=true]{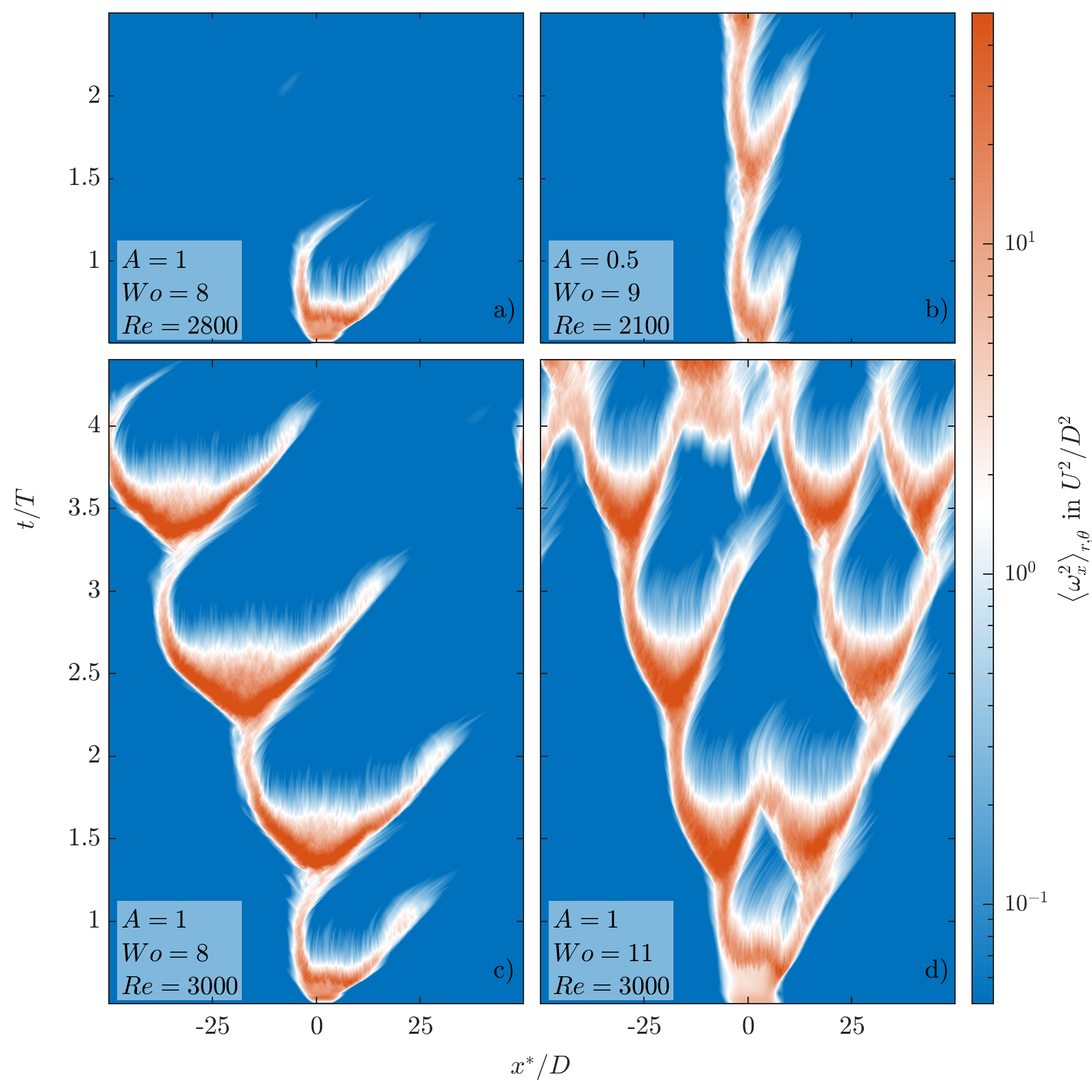}
\caption{Space-time diagrams of the cross section integral of axial vorticity squared $\lla \omega_{x}^{2} \rra_{r,\theta}$ of DNS in a $100D$ long pipe at different flow parameters. The DNS are initialized with the optimal perturbation scaled to $\left|\mathbf{u}^{\prime}_{0}\right|\approx \num{3e-2}$ of magnitude and localized in an axial span of $5D$ following \mod{Feldmann \textit{et al.}}\cite{entropy2021}. The results are shown with respect to a \mod{non-inertial reference frame $x^{*}\left(t\right)=\int_{0}^{t} \bar{U}\left(t'\right) \mathrm{d}t'$, moving with the bulk velocity $\bar{U} \left(t\right)$.}}
\label{fig:fig2}
\end{figure}

\section{DNS Results}\label{sec:DNSresults}
We perform DNS of pulsatile pipe flow at different combinations of $\Reynolds \geq 2100$, $5 \leq \Womersley \leq 19$ and $0.5 \leq \Amplitude \leq 1$. At these flow parameters, turbulence always takes the form of localized turbulent puffs that are modulated by the pulsation, as shown in fig.~\ref{fig:fig2}. See a list of the DNS we analyse here at the end of the manuscript, in tables~\ref{tab:dns} and \ref{tab:MSdns}.

In the parametric space considered here, we observe 4 different behaviors of turbulent puffs. We classify our DNS according to these four behaviors, see tables \ref{tab:dns} and \ref{tab:MSdns}:
\begin{enumerate}
    \item \textbf{First elongation, then rapid decay (RaD)}: the initial helical perturbation, used to initialize the flow, first grows in length and magnitude, and then decays in less than one pulsation period \mod{(figure~\ref{fig:fig2}a)}. \mod{We classify these decay events as deterministic. They are different from decay events that happen (stochastically) after more than one pulsation period, which we classify in another category (3. below).}
    \item \textbf{Localized turbulent structures (Loc)}: the initial helical perturbation localizes in a puff, that is then modulated in length and magnitude by the pulsation and survives for long times without splitting \mod{or decaying}. See an example of this behavior in figure~\ref{fig:fig2}b.
    \item \textbf{Localized structures, then stochastic decay (StD)}: the initial helical perturbation localizes in structures that are modulated, in length and magnitude, by the pulsation. These structures, however, tend to suddenly decay after typically a short number of pulsation periods \mod{(figure~\ref{fig:fig2}c). Although we do not explicitly compute life-time statistics of these cases here, these decay events happen at random times, as in pulsatile pipe flows at lower $\Amplitude \leq 0.4$ \cite{xu2018} or driven with more complex waveforms \cite{moron2022}. In those analyses, the researchers report that turbulent structures decay after a random number of pulsation periods. However, different to SSPF, these decay events are more likely to happen at a particular phase of the period. } 
    \item \textbf{Highly intermittent state (Int)}: the initial helical perturbation localizes in structures modulated by the pulsation. These structures, however, randomly split until the DNS reaches a highly intermittent state where turbulence aggregates in localized structures modulated by the pulsation and separated by laminar patches (figure~\ref{fig:fig2}d). 
\end{enumerate}

In the following section, \S \ref{sec:master-slaveresults}, we study the mechanisms by which puffs are able to survive the pulsation at certain flow parameters. Here we \mod{briefly} analyse the behavior of puffs in pulsatile pipe flow. We focus on two quantities, the upstream front speed of puffs $c_{u}$ and the phase difference $\Delta \phi$ between the volume averaged axial vorticity squared $\lla \omega_{x}^{2} \rra_{r,\theta,x} \left(t \right)$ and the bulk velocity $\bar{U} \left(t\right)$. The former is the main observable, that the results of the EBM are meant to approximate. The latter has to be considered in the EBM to correctly approximate the front speed of puffs, as shown later in \S \ref{sec:ebmresults} and appendix \ref{ap:ebm}.

\begin{figure}[h!]
\centering
\includegraphics[width=\textwidth, trim=0mm 0mm 0mm 0mm, clip=true]{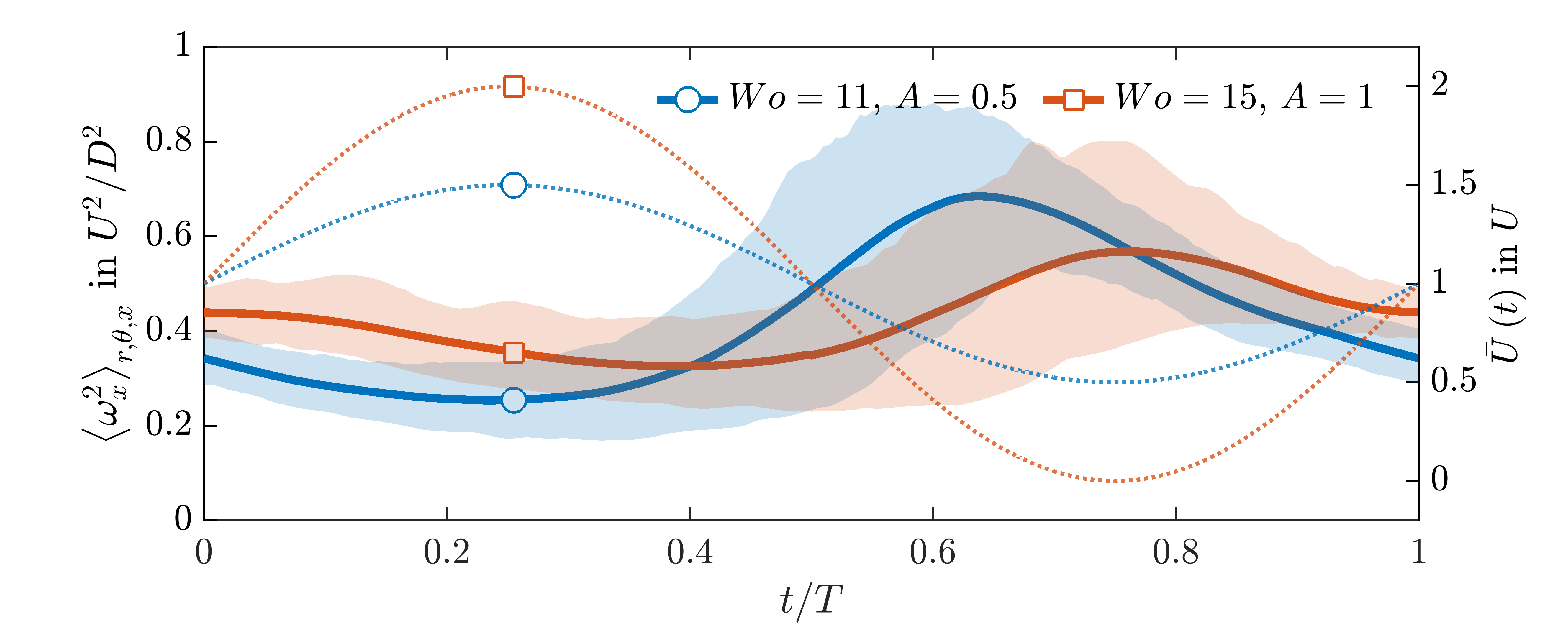}
\caption{\mod{Time profile of $\left \langle \omega_{x}^{2} \right \rangle_{r,\theta,x,t^{*}}$ (solid thick lines) compared with the bulk velocity $\bar{U} \left(t\right)$ (dotted lines); for $\Reynolds=2100$, \Womersley and \Amplitude as indicated in the legend. The shaded regions delimit the volume of data between the max/min $10\%$ percentile of the corresponding $\left \langle \omega_{x}^{2} \right \rangle_{r,\theta,x} \left(t^{*}\right)$ phase-dependent statistics.}}
\label{fig:fig3n}
\end{figure}

\begin{figure}
\centering
\includegraphics[width=\textwidth, trim=0mm 0mm 0mm 0mm, clip=true]{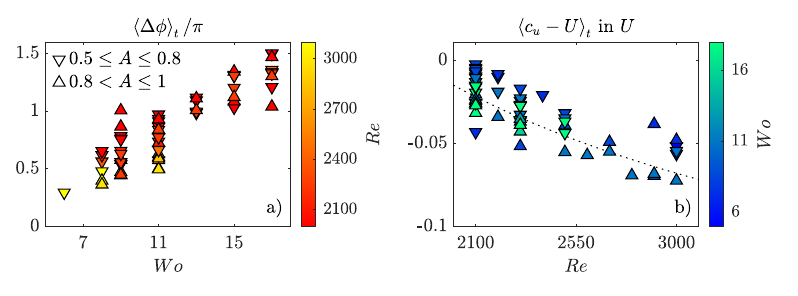}
\caption{Effect of flow parameters on: (a) the phase difference $\Delta \phi$ between bulk velocity $\bar{U}$ and volume-averaged turbulence intensity $\lla \omega_{x}^{2} \rra_{r,\theta,x} \left(t \right)$; (b) upstream front speed $c_{u}$. Each marker corresponds to the time averaged value of either $c_{u}$ or $\Delta \phi$ of an individual DNS of pulsatile pipe flow listed in table~\ref{tab:dns} (and master simulations listed in table~\ref{tab:MSdns}). Downward (upward) pointing triangles correspond to simulations at $0.5\leq \Amplitude \leq 0.8$ $\left( 0.8 < \Amplitude \leq 1\right)$. \mod{The face color indicates \Reynolds in subplot a) and \Womersley in subplot b). The dotted line in subplot b) corresponds to a fit of the upstream front speed of puffs in SSPF: $c_{u}-U \approx 0.28\left[0.024 + \left( \frac{\Reynolds}{1936}\right)^{-0.528} - 1.06\right] $ in $U$, according to  Chen \textit{et al.}  \cite{chen2022}.}}
\label{fig:fig3}
\end{figure}

\mod{We compute the upstream front speed $c_{u}$ by tracking the upstream-most position in the turbulent puff, defined as $x_{u}=\min \left( u \right)$ such that $\lla \omega_{x}^{2} \rra_{r,\theta} \geq \num{1e-1}$ for $x \gtrsim x_{u}$. We observe that the time signal of $\lla \omega_{x}^{2} \rra_{r,\theta,x} \left(t \right)$ approximates a harmonic function, see figure~\ref{fig:fig3n}, at all the flow parameters considered here. Thus, we compute the time-averaged phase difference $\lla \Delta \phi \rra_{t}$ by projecting the time signal of $\lla \omega_{x}^{2} \rra_{r,\theta,x} \left(t \right)$ to a harmonic function, and comparing its phase with the sinusoidal bulk velocity. We repeat this for the data of all the simulations listed in table~\ref{tab:dns} and the master DNS in table~\ref{tab:MSdns}. We show the time averaged upstream front speeds and phase difference of all the simulations together in figure~\ref{fig:fig3}. }

We observe a phase difference between the turbulence intensity (represented by $\lla \omega_{x}^{2} \rra_{r,\theta,x} \left(t \right)$) and bulk velocity consistent with studies of small-to-moderate amplitude $\Amplitude\leq 0.4$ pulsatile pipe flow \citep{xu2018}, and fully turbulent pulsatile channel flows \citep{weng2016}. There, they observed how the phase difference increases for increasing \Womersley, \mod{as we also observe in the cases considered here,} see figure~\ref{fig:fig3}a. \mod{At small $\Reynolds\approx2100$ the phase difference saturates at $\Delta \phi \approx \frac{3\pi}{2}$. There is no apparent effect of \Amplitude on $\Delta \phi$.} 

The upstream front speed, as in the case of SSPF, is mainly affected by \Reynolds, see fig.~\ref{fig:fig3}b. The higher \Reynolds is, the smaller the upstream front speed becomes. According to our results, there is also a weak dependence of the front speed on \Amplitude. The upstream front speed tends to decrease for increasing \Amplitude. \mod{As \Womersley increases, the upstream front speed aproximates the value of SSPF, as turbulence is less affected by the pulsation.}

\modd{\section{Master--slave causal analysis}\label{sec:master-slaveresults}}
\begin{figure}
\centering
\includegraphics[width=\textwidth, trim=0mm 0mm 0mm 0mm, clip=true]{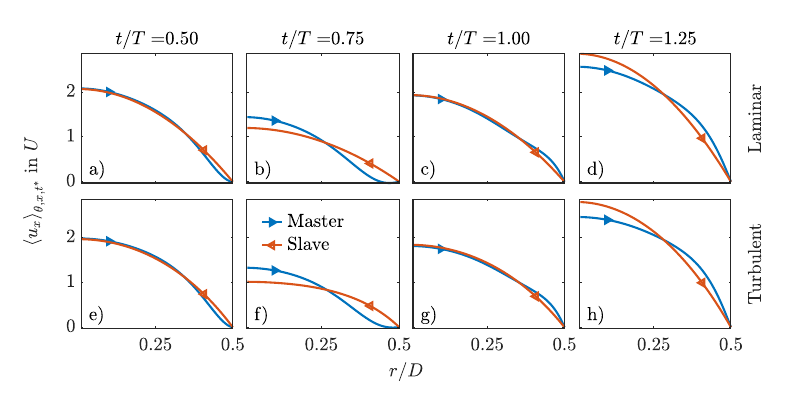}
\caption{Phase $\left( t^{*} \right)$ averaged mean profiles of master (M) and slave (S) simulations at $\Reynolds=2100$, $\Womersley=11$, $\Amplitude=0.5$, \mod{for the case of fully laminar (a,b,c,d) and turbulent flows (e,f,g,h)}. The phase of the period is indicated at the top of each subplot column.}
\label{fig:fig1}
\end{figure}

\begin{figure}
\centering
\includegraphics[width=\textwidth, trim=0mm 0mm 0mm 0mm, clip=true]{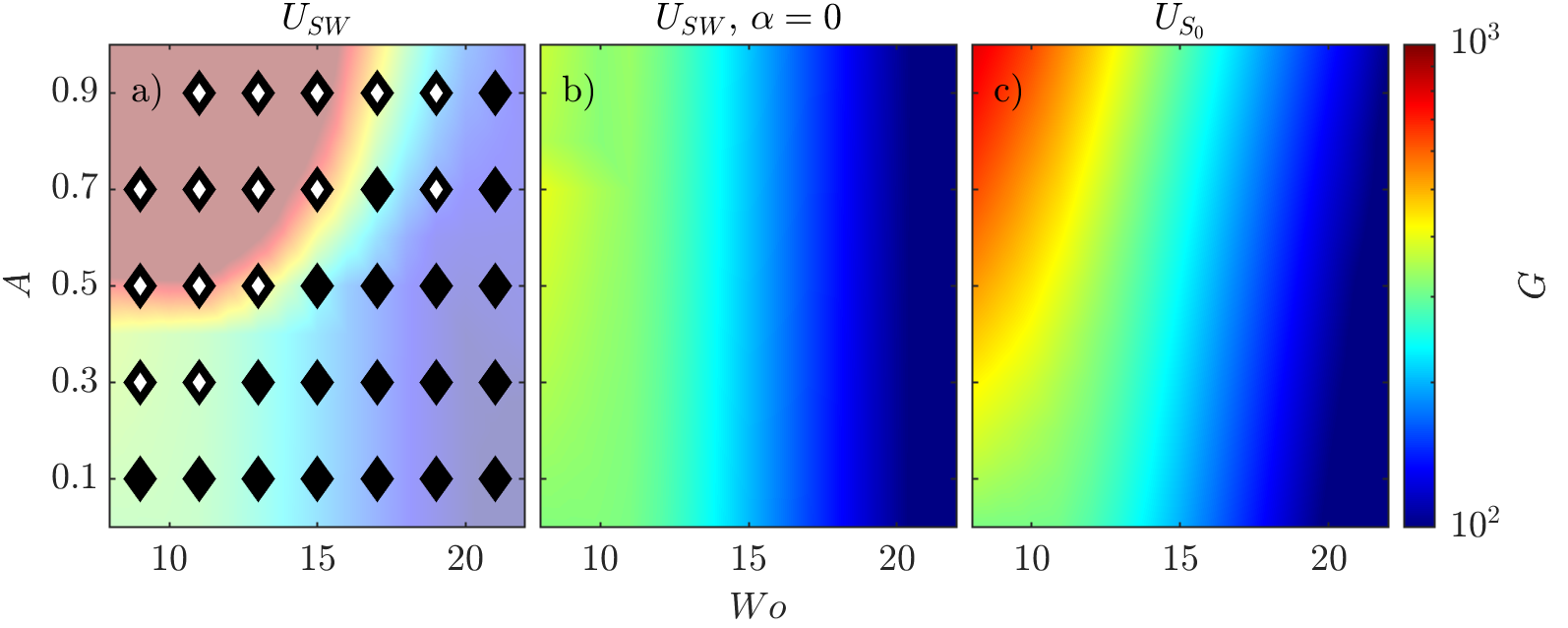}
\caption{With colors, the maximum transient energy growth of perturbations on top of master $U_{SW}$ (a,b) and slave $U_{S}$ (c) laminar profiles. Note that the color scale is limited to a maximum of $G \leq 10^{3}$ for clarity. The results correspond to different $\Womersley$ and $\Amplitude$, at $\Reynolds=2100$. In a) the maximum energy growth of any helical or stream-wise constant perturbation with $\alpha \geq 0$ and $m=1$ on top of $U_{SW}$ profiles. The symbols correspond to DNS results in a $L=100D$ long pipe initialized with a localized turbulent puff. Symbols denote pairs of DNS whose master simulation has puffs that survive for long times. Hollow symbols indicate pairs of DNS whose slave simulation has puffs that decay at $t\leq t_{d}= 120 D/U$, \mod{while filled symbols indicate slave DNS where puffs survive for longer times}. In b), the maximum energy growth of streamwise-constant perturbations with axial $\alpha=0$ and azimuthal $m=1$ wavenumbers on top of the $U_{SW}$ profile. In c) the maximum energy growth of any perturbation $\alpha \geq 0$ and $m=1$ on top of the \mod{laminar slave} $U_{S}$ profile. }
\label{fig:fig6}
\end{figure}

\begin{figure}
\centering
\includegraphics[width=\textwidth, trim=0mm 0mm 0mm 0mm, clip=true]{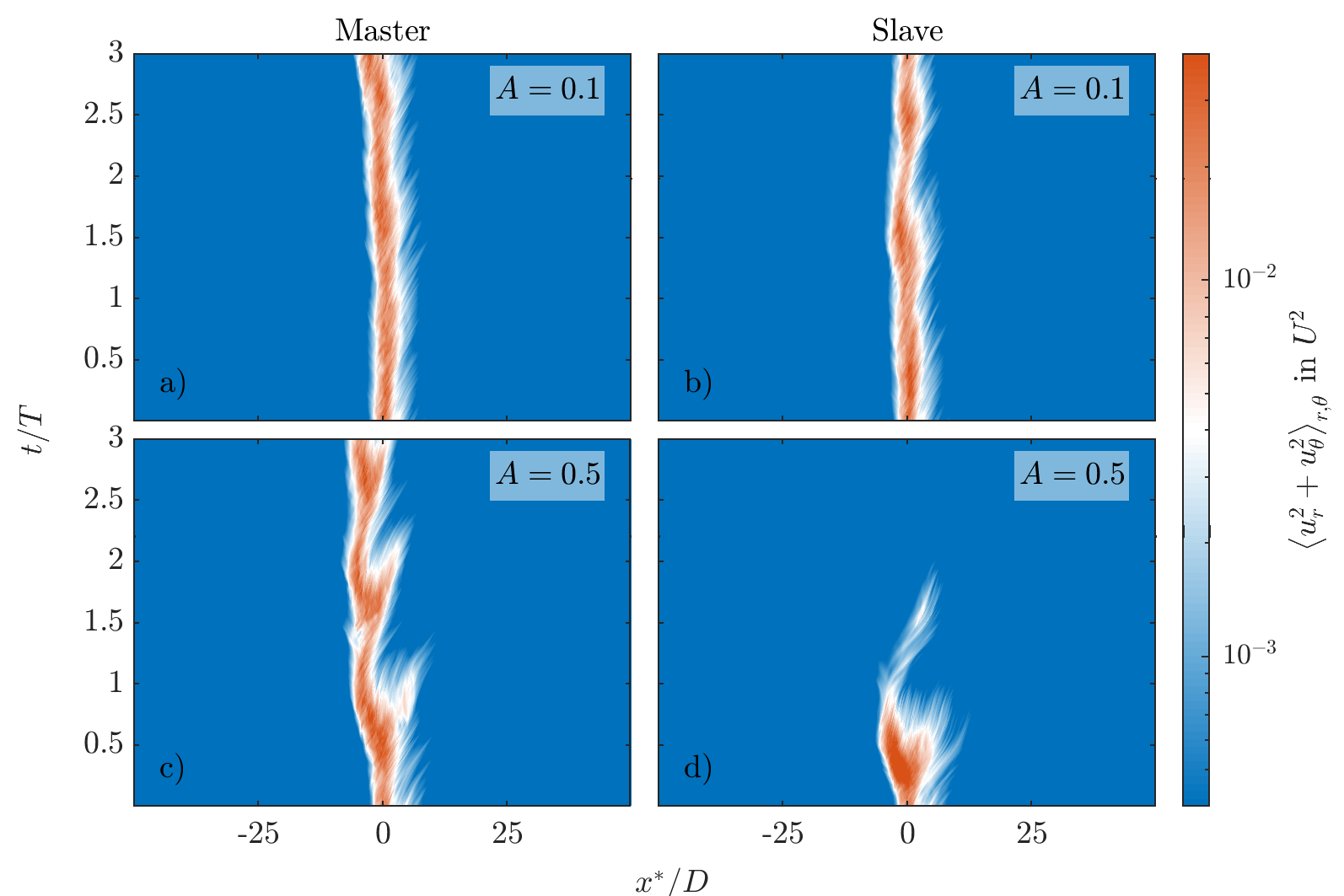}
\caption{Space-time diagrams of the cross section integral of turbulent cross section kinetic energy, in master-slave DNS at $\Reynolds=2100$ and $\Womersley=11$, and two different \Amplitude. The results correspond to two pairs of master-slave DNS in a $L=100D$ long pipe initialized with a localized turbulent puff. The results are plotted with respect to a moving reference frame moving at the bulk velocity $\bar{U}$. a) and c) correspond to master simulations, b) and d) to slave simulations.}
\label{fig:fig7}
\end{figure}

\modd{Our hypothesis is that, as in SSPF \cite{barkley2015rise,Song2017,chen2022}, the behavior of puffs in pulsatile pipe flow is determined by the shape of the axial velocity profile $\lla u_{x} \rra_{\theta} \left(x=x_{u},r,t\right)$ at the upstream front $x \approx x_{u}$ of the puff. In the ideal case of an extremely long pipe, and a single localized turbulent puff, the mean profile in the pipe $\lla u_{x} \rra_{\theta,x} \left(r,t\right)$ approximates this profile. The idea is that puffs take advantage of two characteristics of the mean (upstream) velocity profile. One is the mean shear, as in SSPF, that, for the pulsatile case, is time-dependent. The other is the instability of the SW profile, that is linked with the presence of inflection points. We perform causal analysis of pulsatile pipe flow with the goal to separate the two mechanisms. Inspired by \citet{tuerke_jimenez_2013}, we perform DNS of pulsatile pipe flow with prescribed mean profiles in which inflection points are erased.}

\modd{Following \citet{vela-martin_2021} and references therein, we perform pairs of DNS that run in parallel to each other. In each pair, one of the simulations is a full DNS of pulsatile pipe flow, referred to as the master. The other simulation, referred to as the slave, uses the information of the master to modify some of its characteristics. Specifically, the slave simulation uses the instantaneous mean profile of the master $U_{M}=\lla u_{x} \rra_{\theta,x} \left(r,t\right)$ to generate its instantaneous mean profile $U_{S} \left(r,t\right)$. A more detailed explanation of the methods used to compute $U_{S} \left(r,t\right)$ is given in the appendix \ref{ap:master_slave}.}

\modd{We design time-dependent artificial (slave) profiles that, in the laminar case, have a shear that monotonically decreases from the center-line to the pipe wall. Additionally, at each time step, the slave profiles have the same kinetic energy as the corresponding SW profile. See an example of the slave laminar profiles at $\Reynolds=2100$, $\Womersley=11$ and $\Amplitude=0.5$ in figure~\ref{fig:fig1}(a,b,c,d). The slave profiles $U_{S}$ have a similar magnitude to the actual SW profile, they are also time dependent but, different to the SW profile, they do not have inflection points. }

\modd{\subsection{Master--slave TGA}}
\modd{According to our hypothesis, the absence of inflection points prevents the outstanding transient growth of helical perturbations \cite{moron2022}. We test this hypothesis by performing TGA on the slave laminar profiles at many flow parameters, and comparing the results with TGA on the corresponding SW profile. Note that, in the laminar case, the mean profile of the master is equal to the SW profile.}

\modd{The TGA returns the shape (the radial profile and the axial $\alpha$ and azimuthal $m$ wavenumber) of the infinitesimal perturbations that reach the highest energy growth $G$ on top of a given flow profile. The TGA also returns the optimal time $t_{0}$ to trigger the perturbation, and the final time $t_{f}$ where perturbations reach the highest energy growth $G$. In this paper, we do not show the radial shape of the optimal perturbations, as our main interest is to check whether the slave profiles are highly susceptible to the growth of helical perturbations with $\alpha >0$ and $m=1$, or not. We also do not analyze $t_{f}$, neither the optimal time to trigger the perturbations, which is always around $t_{0}\approx T/2$ \cite{xu2021non}. We report the maximum transient growth we observe at any point in time during one pulsation period $t_{f}-t_{0} \leq T$. We plot as colormaps the results of our TGA in figure~\ref{fig:fig6}.}

\modd{We perform two different TGA of our laminar (SW) master profiles. In the first TGA, we consider all possible perturbations with $\alpha \geq 0$ and $m=1$. At several flow parameters, the energy growth of helical perturbations, with $\alpha>0$, is much higher than for any other perturbation \cite{xu2021non}. The flow parameters at which this happens can be clearly identified as the intense red region in figure~\ref{fig:fig6}a. In the second TGA, we fix $\alpha=0$ and $m=1$, and compute the growth of streamwise-constant perturbations. According to our TGA, and as observed by Xu \textit{et al.} \cite{xu2021non}, the SW profile is susceptible to the growth of streamwise constant perturbations, see figure~\ref{fig:fig6}b. Only at high \Womersley or small \Amplitude, the growth of streamwise-constant perturbations is larger than the growth of helical perturbations.}

\modd{For slave profiles we perform just one TGA, in which we consider all possible perturbations with $\alpha \geq 0$ and $m=1$. The results of this analysis are shown in figure~\ref{fig:fig6}c. At all the flow parameters we consider here, the perturbations that reach the highest energy growth are streamwise-constant perturbations (with $\alpha=0$ and $m=1$). Note that their energy growth is similar to the ones obtained in the SW profile, compare figure~\ref{fig:fig6}c and b. This confirms that our slave profiles qualitatively capture the transient growth characteristics of streamwise constant perturbations in the SW profile. This growth is linked to the mean shear of the flow, as in SSPF, and not to the presence of inflection points \cite{moron2022}. It also shows that, unlike the actual SW profile, our slave profiles are not susceptible to the outstanding growth of helical perturbations.}

\modd{\subsection{Master--slave DNS}}
\modd{We perform pairs of master-slave DNS at several flow parameters. In each pair, both simulations are initialized with the same turbulent field, that has a (single) turbulent structure. The slave profiles have a shear that monotonically decreases from the center-line to the pipe wall. However, by construction, the kinetic energy of the slave profiles matches the kinetic energy of the instantaneous master profile. The slave profiles are time dependent and do not have inflection points. They are also more blunted, the more deformed the corresponding master profiles are with respect to the SW profile, see appendix \ref{ap:master_slave}.}

\modd{Find an example of the slave profiles, and their master counterparts, in DNS at $\Reynolds=2100$, $\Womersley=11$ and $\Amplitude=0.5$ in figure~\ref{fig:fig1}(e,f,g,h). For this parameter set, the $U_{SW}$ profile is susceptible to the growth of helical perturbations, as seen in fig.~\ref{fig:fig6}a, and, while the master simulation has a localized puff that survives for a long time, fig.~\ref{fig:fig7}c, the puff in the slave simulation decays in less than two pulsation periods, fig.~\ref{fig:fig7}d. At $\Amplitude=0.1$ and $\Womersley=11$, see fig.~\ref{fig:fig7}a and b, both the slave and master profiles are not susceptible to the growth of helical perturbations, fig.~\ref{fig:fig6}. At these parameters, puffs survive during the whole simulation time for both, slave and master simulations.}

\modd{In order to check whether this behavior is reproduced at other flow parameters, we set an heuristic time threshold at $t_{d} = 120 D/U$, that corresponds to a time span of more than 3 pulsation periods for all cases considered here $t_{d}>3T$. Our master-slave DNS are classified according to whether the slave simulation shows a decay event or not. See in figure \ref{fig:fig6}a a graphic representation of this classification for different combinations of \Amplitude and \Womersley, at $\Reynolds =2100$. Full symbols denote slave simulation in which the puff survives for long times. Hollow symbols indicate slave simulations that show puff decay at $t\leq t_{d}$. In all the cases the master simulation survives. There is a clear boundary between cases that show quick puff decay in the slave simulations and those which do not. At $\Amplitude \gtrsim 0.5$ and $8 \lesssim \Womersley \lesssim 17$ pulsatile pipe flows are highly susceptible to the growth of helical perturbations, fig.~\ref{fig:fig6}a. As seen in fig.~\ref{fig:fig6}a, at these flow parameters, after suppressing the inflection points, puffs quickly decay in the corresponding slave DNS. At $\Amplitude \lesssim 0.3$ and/or $\Womersley \gtrsim 20$, the growth of the helical perturbations is smaller than that of the streamwise constant perturbations, see fig.~\ref{fig:fig6}a compared to \ref{fig:fig6}b. At these flow parameters, eliminating the inflection points has no effect on the lifetime of turbulent puffs. This result further supports that, at certain \Reynolds, \Womersley and \Amplitude, as soon as puffs cannot make use of the inflection points to survive, they quickly decay.}

\begin{figure}
\centering
\includegraphics[width=0.95\textwidth, trim=0mm 0mm 0mm 0mm, clip=true]{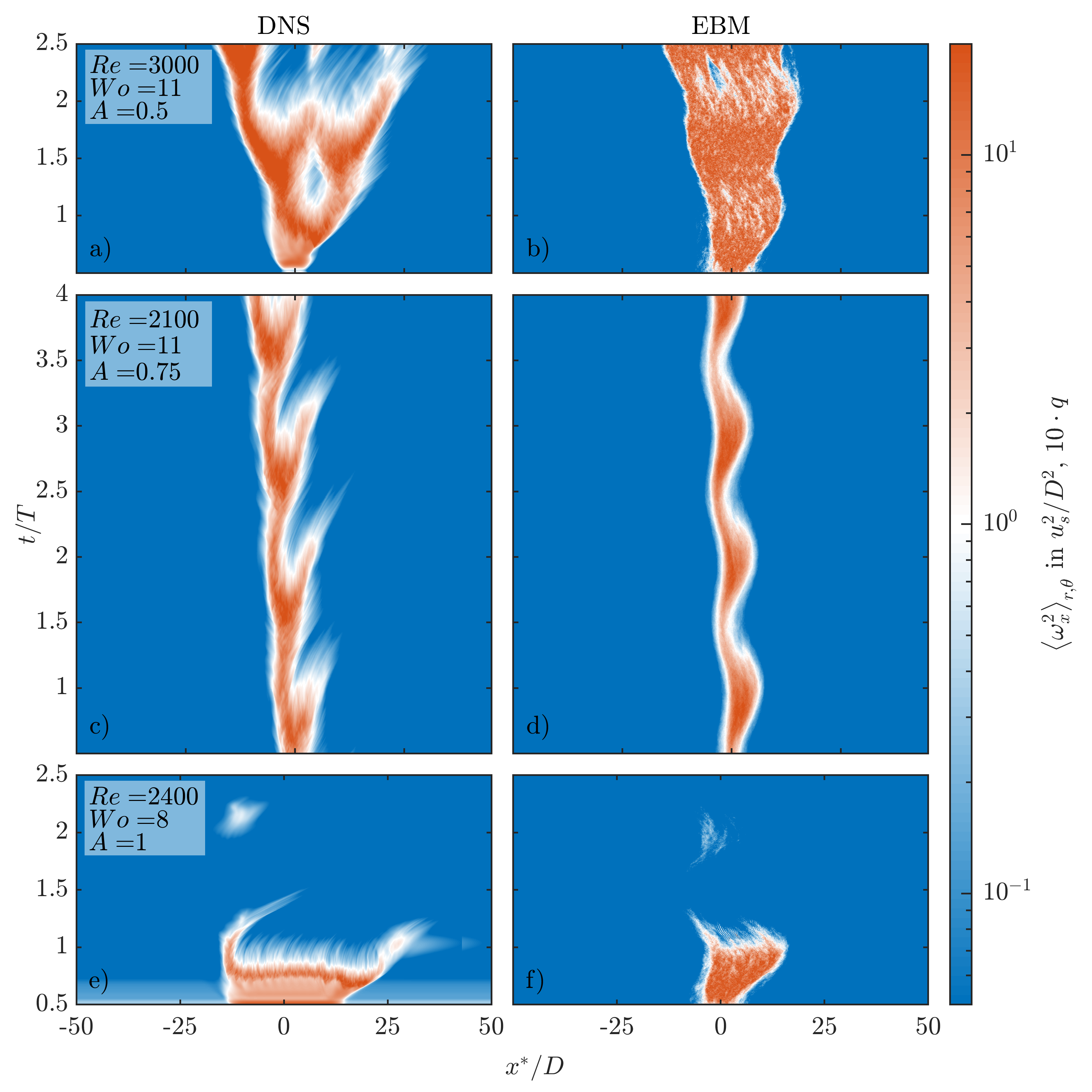}
\caption{Space-time diagrams of the cross section integral of axial vorticity squared of DNS (left plots: a, c, e) and $10 \cdot q$ of the EBM (right plots b, d, f) in a $100D$ long pipe. The DNS are initialized with the optimal perturbation scaled to $\left|\mathbf{u}^{\prime}_{0}\right|\approx \num{3e-2}$ of magnitude and localized in a span of $5D$ as in \cite{entropy2021}, and the model simulations with a localized perturbation of length $5D$. The figure is presented with respect to a moving frame $x^{*}$, moving with the bulk velocity $\bar{U}$. a) and b) correspond to $\Reynolds=3000$, $\Womersley=11$, $\Amplitude=0.5$. c) and d) correspond to $\Reynolds=2100$, $\Womersley=11$, $\Amplitude=0.75$. e) and f) correspond to $\Reynolds=2400$, $\Womersley=8$, $\Amplitude=1$. This last result is reproduced from \citet{entropy2021}.}
\label{fig:fig8}
\end{figure}

\section{EBM results}\label{sec:ebmresults}
In this section we compare the DNS results with the EBM. The EBM includes two main changes compared to the BM. These changes are inspired by our analyses in the previous sections, and model the two turbulence production mechanisms discussed in this study. One is the phase-lagged effect of a time-varying mean shear, and, \mod{the other, the linear instabilities due to the inflection points in the laminar profile.} Find a detailed description of the EBM in appendix~\ref{ap:ebm}. 

\begin{figure}
\centering
\includegraphics[width=\textwidth, trim=0mm 0mm 0mm 0mm, clip=true]{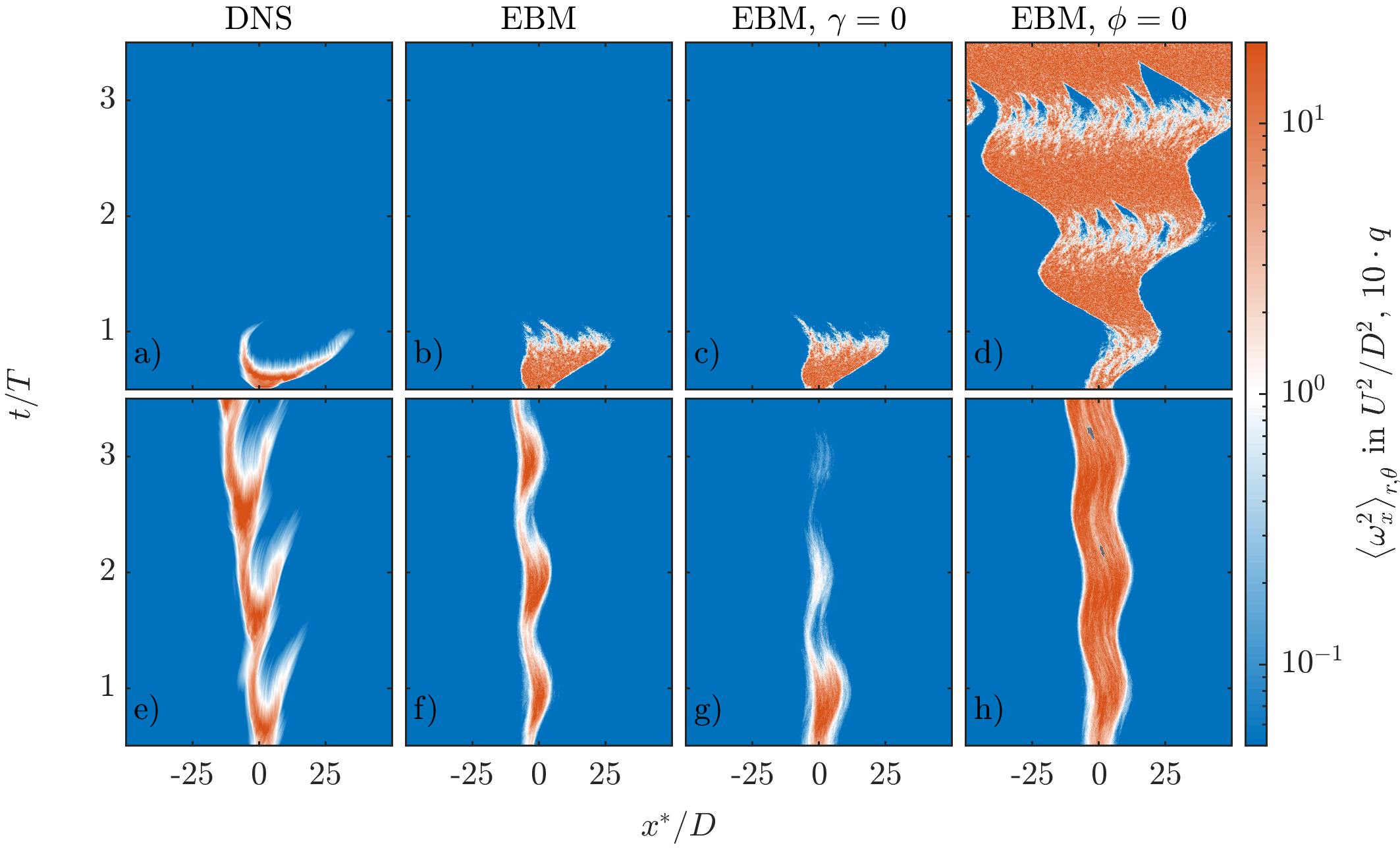}
\caption{Space-time diagrams of the cross section integral of axial vorticity squared of DNS (left plots: a, e) and $10 \cdot q$ of the EBM (b, c, d, f, g, h). The results correspond to DNS and model simulations in a $100D$ long pipe at $\Amplitude=1$. The top panels (a,b,c,d) at $\Reynolds=3000$ and $\Womersley=6$. The bottom panels at $\Reynolds=2200$ and $\Womersley=11$. The DNS (a,e) are initialized with the optimal perturbation scaled to $\left|\mathbf{u}^{\prime}_{0}\right|\approx \num{3e-2}$ of magnitude and localized in a span of $5D$ following \cite{entropy2021}, while the model simulations with a localized perturbation of length $5D$. The figure is presented with respect to a moving frame $x^{*}$, moving with the bulk velocity $\bar{U}$. Panels b, f correspond to EBM simulations with the \mod{fitted} parameters listed in table~\ref{tab:params}. Panels c, g correspond to EBM simulations with the parameters listed in table~\ref{tab:params} but $\gamma=0$. Panels d, h correspond to EBM simulations with the parameters listed in table~\ref{tab:params} but $\phi=0$.}
\label{fig:fig9}
\end{figure}

\begin{figure}
\centering
\includegraphics[width=\textwidth, trim=0mm 0mm 0mm 0mm, clip=true]{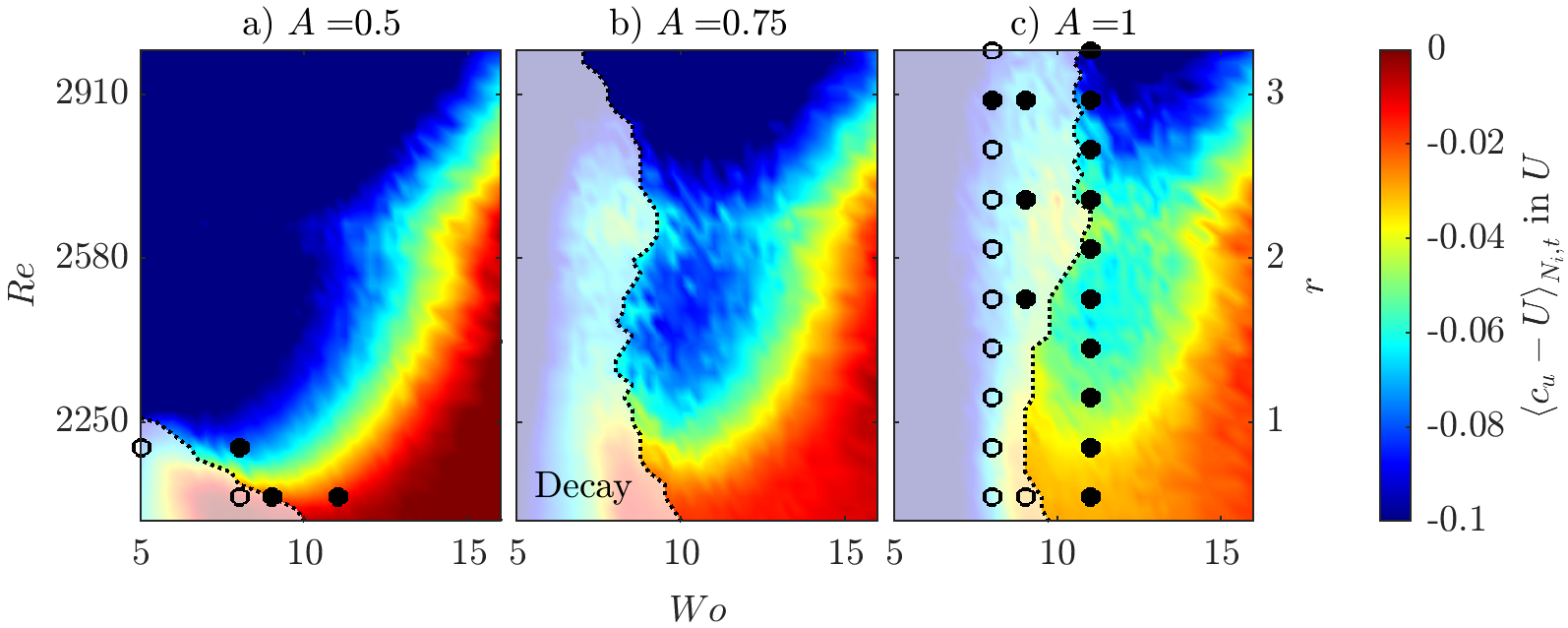}
\caption{With colors, time \mod{$\left(t\right)$} and ensemble \mod{$\left(N_{i}\right)$} averaged upstream front speed $c_{u}$ of $q$ according to simulations of the EBM at several \Reynolds, \Womersley and at three different pulsation amplitudes: in a) at $\Amplitude=0.5$, in b) at $\Amplitude=0.75$ and in c) at $\Amplitude=1$. Each panel corresponds to the interpolated results of an original set of $39\times 37$ \Reynolds and \Womersley combinations. Each combination of \Reynolds and \Womersley has $N_{i}=50$ individual simulations. The dotted lines denote the threshold between \Reynolds and \Womersley cases where more than half of the $N_{i}$ simulations show turbulence decay, $\lla q \rra_{x} \leq 0.01$, before $t/T<4$. With black dots we denote the observed survive/decay behaviors of different DNS at parameters close to the decay threshold. Filled dots correspond to simulations where turbulence survives for long times, or splits/elongates. Hollow points denote DNS where we observe turbulence decay at $t/T\leq 8$.}
\label{fig:fig10}
\end{figure}

In figure~\ref{fig:fig8}, we include three examples of DNS and EBM results comparisons. The model is able to capture reasonably well the turbulent front speed and turbulence behavior of all the cases, as seen qualitatively in figure~\ref{fig:fig8}. With only some exceptions, we observe a good agreement between the EBM and all the DNS listed in tables \ref{tab:dns} and \mod{(only master DNS)}~\ref{tab:MSdns}. 

\subsection{Turbulence production in the EBM}
In the EBM, there are two main sources of turbulence intensity $q$ production, see eq.~\ref{eq:loc_q_EBM}. One is the time varying \Reynolds number that is modelled with the term $r \bar{U}\left( t+\phi \right)$, being $\phi$ a phase lag that only depends on \Womersley. The second one is the instability of the mean profile that is modelled by the product $\gamma \lambda \left(t \right)$. Here $\lambda \left(t \right)$ is the growth rate of the instantaneous instability of the laminar SW profile, as computed by \citet{moron2022} \mod{using a numerical method first developed by Meseguer \textit{et al.} \cite{Meseguer}}. The parameter $\gamma$ \mod{only depends on \Reynolds and \Womersley}. Here we test the effect of ignoring these two mechanisms, one at a time. In figure~\ref{fig:fig9} we show further comparisons between DNS and EBM results. We show results of the EBM model with the fitted parameters listed in table~\ref{tab:params} and also with either $\phi=0$ or $\gamma=0$ \mod{that is without phase lag or without inflectional instability.}

At $\phi=0$ the model results clearly \mod{differ} from the DNS ones at most flow parameters, see fig.~\ref{fig:fig9}d and h. For instance, at $\Reynolds=3000$, $\Womersley=6$ and $\Amplitude=1$, our DNS shows rapid decaying puffs (RaD), see fig.~\ref{fig:fig9}a. However, at \mod{$\phi=0$}, the EBM returns an ever-elongating puff, see fig.~\ref{fig:fig9}d. Moreover at $\Reynolds=2200$, $\Womersley=11$ and $\Amplitude=1$, puffs \mod{are highly modulated by the pulsation in the DNS, see fig.~\ref{fig:fig9}e. However, if one sets $\phi=0$ in the EBM the behavior of the puff is less affected by the pulsation, see fig.~\ref{fig:fig9}h.} Although we do not show it here, we see further divergences between model and DNS results at additional flow parameters when $\phi=0$. This confirms that turbulence perceives the pulsation at a certain \mod{\Womersley-dependent} phase lag with respect to the bulk velocity, as seen in fig.~\ref{fig:fig3}a, and also observed by \citet{weng2016} and \mod{other} studies. 

We also show results of the EBM with $\gamma=0$, see fig.~\ref{fig:fig9}c and g. At $\Reynolds=3000$, $\Womersley=6$ and $\Amplitude=1$, there is no apparent effect of ignoring the linear instability of the mean profile, see fig.~\ref{fig:fig9}a and c. At these flow parameters turbulence rapidly decays due to the effect of the pulsation, regardless of the presence of inflection points in the profile. However, at $\Reynolds=2200$, $\Womersley=11$ and $\Amplitude=1$, while puffs remain localized in the DNS, see fig.~\ref{fig:fig9}e, at $\gamma=0$, puffs in the EBM quickly decay, see fig.~\ref{fig:fig9}g. This is similar to what we observe in our master-slave DNS shown in fig.~\ref{fig:fig6}c. At these flow parameters, without the inflection points and their corresponding linear instability, puffs quickly decay due to the effect of the pulsation.

According to our results with the EBM, one needs to include the phase lag $\phi$ and the effect of the instantaneous instability in the model, to correctly approximate the behavior of puffs in pulsatile pipe flow.

\subsection{Parametric study using the EBM}
We use the EBM to perform simulations in a large \Womersley and \Reynolds parametric space at three different amplitudes, $\Amplitude=0.5$, $\Amplitude=0.75$ and $\Amplitude=1$. At each $\Amplitude$, we consider $39$ equispaced \Reynolds values between $2050\leq \Reynolds \leq 3000$ and $37$ \Womersley values between $5\leq \Womersley \leq 16$. For each combination of \Womersley, \Reynolds and \Amplitude we perform $N_{i}=50$ EBM simulations for $t/T\leq4$, and compute the time and ensemble averaged upstream front speed of turbulent patches, $\lla c_{u} - U\rra_{N_{i},t}$. We stop the simulations either at $t/T=4$ or when turbulence decays in the whole domain, $\lla q \rra_{x} \leq 0.01$. 

\mod{We show the ensemble and time-averaged upstream front speed of the EBM as a colormap in figure~\ref{fig:fig10}}, together with an empiric threshold for $q$ decay. The threshold separates the parametric regions where more than half of the $N_{i}$ \mod{EBM} simulations decay, at $t/T \leq 4$, from the rest. It can be understood as a critical $\Reynolds_{c} \left(\Womersley,\Amplitude\right)$, up to which all \mod{modelled} puffs are more likely to decay than survive. This critical $\Reynolds_{c}$ highly depends on $\Womersley$ and $\Amplitude$. At low $\Womersley$, the system \mod{behaves more quasi-steadily} and puffs need a higher mean $\Reynolds$ to survive the phases of the period where $\bar{U}\left(t+\phi\right)<U$. The amplitude $\Amplitude$ sets the minimum $\Reynolds$ at each $\Womersley$ \citep{xu2017}. As \Amplitude increases, this minimum \Reynolds increases. At $\Amplitude=1$, independently of the selected $\Reynolds \leq 3000$, puffs show a high chance to decay as long as $\Womersley \leq 10$. 

Regarding the upstream front speed, at high $\Womersley$ and independently of the selected $\Amplitude$, $\lla c_{u} - U\rra_{N_{i},t}$ decreases as $\Reynolds$ increases. This is expected as the behavior of puffs at high $\Womersley$ is similar to the behavior of puffs in SSPF and the upstream front speed of puffs in SSPF decreases for increasing \Reynolds \cite{barkley2015rise}. As $\Womersley$ decreases, at $10 \lesssim \Womersley \lesssim 16$, $\lla c_{u} - U\rra_{N_{i},t}$ decreases. At these frequencies, $\gamma \lambda >0$ for some phases of the period, which increases the $q$ production. This effect, can be understood as an increase of \Reynolds, and therefore causes a lower averaged $c_{u}$. As $\Amplitude$ (and/or $\Reynolds$) increases, $\lambda$ increases \citep{moron2022}, which results in a lower upstream front speed for all the parameters considered here. At $\Womersley \approx 10$ the system is close to the decay threshold discussed above. Puffs tend to accelerate as they decay, which explains the increase of $\lla c_{u} - U\rra_{N_{i},t}$ at these pulsation frequencies. 

However our results show that, as $\Womersley$ further decreases, $\lla c_{u} - U\rra_{N_{i},t}$ decreases again. This is due to the way we initialize the EBM simulations. At $t_{0}=T/2$ puffs tend to elongate since $r \bar{U} \left(t+\phi\right) \geq 0$ and $\gamma \lambda \geq 0$. Therefore, during the initial phase of the EBM simulations, at $\Womersley \lesssim 8$ puffs first rapidly elongate $\left( c_{u} - U< 0 \right)$, and then quickly decay when $\bar{U}\left(t+\phi\right)<U$. Since the decay  \mod{happens} faster than the elongation, the averaged front speed is $\lla c_{u}-U<0 \rra_{N_{i},t}$.
 
\modd{\subsection{Assessment of the EBM results}
The EBM qualitatively captures the behavior of the upstream front speed observed in the DNS, as seen after comparing figure~\ref{fig:fig10} with figure~\ref{fig:fig3}b. As in the DNS, the upstream front speed in the EBM decreases for increasing \Reynolds and \Amplitude, and, at high \Womersley, it approaches the values of the upstream front speed of puffs in SSPF. The quantitative values of the upstream front speed of model and DNS are also similar.

In figure~\ref{fig:fig10}c, additionally we represent with symbols the survival/decay behaviors of the DNS listed in table~\ref{tab:dns}, whose flow parameters are close to the decay thresholds of the EBM. Hollow symbols represent DNS where we observe puff decay at $t/T \leq 8$, and solid ones, DNS that show turbulence survival for long times. The model approximates relatively well the minimum \Womersley at each \Reynolds and \Amplitude, where turbulence does not decay after a short number of pulsation periods. The match between EBM and DNS results is better at smaller \Amplitude, like $\Amplitude=0.5$ than at higher \Amplitude. At $\Amplitude=1$, while the model has a threshold to decay close to $\Womersley=10$, in the DNS the threshold seems to be closer to $\Womersley=8$, and slightly change as \Reynolds increases.

The observed discrepancies between model and DNS results are rooted in the limitations of the EBM, and the underlying BM, which are described in detail in \S \ref{sec:limitsEBM}.} 

\section{Conclusions}\label{sec:conclusion}
In this paper we study the behavior of turbulence in transitional pulsatile pipe flow at $2100 \leq \Reynolds \leq 3000$, $5 \leq \Womersley \leq 20$ and $0.5 \leq \Amplitude \leq 1$. At these flow parameters, turbulence tends to first appear in the form of localized turbulent structures whose length and magnitude are modulated by the pulsation. We perform $71$ DNS, and identify different behaviors of these turbulent puffs. At some flow parameters the structures decay after one pulsation period, or after a short number of periods, but always at a fixed phase of the pulsation. At other flow parameters, they survive the pulsation and remain localized for asymptotically long times or randomly split until the flow reaches a highly intermittent state with localized turbulent patches that split/decay in a quiescent laminar flow. 


By performing a causal analysis we show that, at certain flow parameters, these patches actively make use of the inflection points in the quiescent laminar profile, and their corresponding instabilities, to survive the pulsation.

Using the lessons learned from the DNS results and the results of our causal analysis, we adapt the BM to pulsatile pipe flow. The new EBM is able to qualitatively approximate the behavior of turbulent patches at all studied flow parameters. Specifically it reproduces reasonably well the behavior of turbulent front speeds in pulsatile pipe flow, including their dependence on \Reynolds, \Amplitude and \Womersley. It also approximates the thresholds of rapid turbulence decay in terms of \Reynolds, \Amplitude and \Womersley. 

According to our DNS, causal analysis and the EBM results, turbulence in pulsatile pipe flow at these flow parameters makes use of mainly two mechanisms to survive. The first mechanism is the turbulent production due to the mean shear, which is maximum in the phases of the period where the bulk velocity $\bar{U}\left(t+\phi\right)>U$. This production has a certain lag $\phi$ with respect to the pulsation, which is mainly set by \Womersley. The second mechanism is the instantaneous instability of the quiescent laminar SW flow. As long as puffs remain localized in pulsatile pipe flow, and surrounded by a quiescent laminar profile with inflection points, they can take advantage of this instability to increase the turbulent production at certain phases of the period. 

As future steps, one could extend the analysis to higher \Reynolds and different pulsation waveforms. We expect that, as \Reynolds increases, non-linear effects become more important, as reported by \citet{pier_schmid_2017}, in detriment to the mean shear and the effect of inflection points. In fact, as the turbulent fraction increases at high \Reynolds, the inflection points should play a smaller role in turbulence production. Also at higher \Reynolds recent studies show that the waveform of the pulsation has a big impact on the behavior of the system, \citep{waveforms2018,Lopez2023}. It would be interesting to determine the parametric thresholds where the behavior of turbulence in pulsatile pipe flow transitions from localised modulated puffs, to fully turbulent, but statistically phase dependent, flow.

\section*{Acknowledgments}
This work was funded by the German Research Foundation (DFG) through the research unit \pulsfor under grant \texttt{AV~120/6-1}, which is gratefully acknowledged. Computational resources were partially provided by the North German Supercomputing Alliance (HLRN), which are also gratefully acknowledged. The authors also appreciate many fruitful and inspiring discussions with the \texttt{fluid modeling and simulation} group at ZARM, in particular Dr. Feldmann and Dr. Vela-Martin for suggesting the use of a phase lag in the EBM. We would also like to acknowledge Jose Cela París for inspiring the small variations formulation we use to compute our slave profiles. Finally, the authors are grateful to Prof. Dr. Dwight Barkley, for fruitful discussions and suggestions to improve the manuscript.

\nocite{*}
\bibliography{Foil}


\begin{table}[h!] 
\centering
\begin{tabular}{c|ccc|ccc|ccccc|cc} 
Case & \Reynolds & \Womersley & \Amplitude & $N_{r}$ & $N_{\theta}$ & $N_{x}$ & $\Reynolds_{\tau}$ & $\Delta r_{\min}^{+}$ & $\Delta r_{\max}^{+}$ & $\Delta R \theta^{+}$ & $\Delta x^{+}$ & $NT$ & Behavior \\ \hline 
1 & 2100 & 8 & 0.50 & 96 & 80 & 1200 & 96.49 & 0.022 & 1.47 & 2.53 & 5.36 & 8.3 & StD\\ 
2 & 2100 & 9 & 0.50 & 96 & 96 & 1600 & 99.52 & 0.023 & 1.52 & 2.17 & 4.15 & 12.4 & Loc\\ 
3 & 2100 & 9 & 1.00 & 96 & 96 & 1536 & 123.55 & 0.028 & 1.89 & 2.70 & 5.36 & 23.0 & StD\\ 
4 & 2100 & 11 & 0.50 & 96 & 80 & 1800 & 104.68 & 0.024 & 1.60 & 2.74 & 3.88 & 188.0 & Loc\\ 
5 & 2100 & 11 & 0.75 & 96 & 80 & 1800 & 118.75 & 0.027 & 1.81 & 3.11 & 4.40 & 7.9 & Loc\\ 
6 & 2100 & 11 & 1.00 & 96 & 80 & 1800 & 131.87 & 0.030 & 2.01 & 3.45 & 4.88 & 193.0 & Loc\\ 
7 & 2100 & 15 & 0.50 & 96 & 80 & 1800 & 114.92 & 0.026 & 1.75 & 3.01 & 4.26 & 10.0 & Loc\\ 
8 & 2100 & 15 & 1.00 & 96 & 80 & 1800 & 146.73 & 0.034 & 2.24 & 3.84 & 5.43 & \mod{64.0} & Loc\\ 
9 & 2100 & 17 & 0.50 & 96 & 80 & 1800 & 118.95 & 0.027 & 1.82 & 3.11 & 4.41 & 10.0 & Loc\\ 
10 & 2100 & 17 & 1.00 & 96 & 80 & 1800 & 153.99 & 0.035 & 2.35 & 4.03 & 5.70 & 10.0 & Loc\\ 
11 & 2200 & 5 & 0.50 & 96 & 80 & 1200 & 88.74 & 0.020 & 1.35 & 2.32 & 4.93 & 3.2 & RaD\\ 
12 & 2200 & 8 & 0.50 & 96 & 80 & 1200 & 98.89 & 0.023 & 1.51 & 2.59 & 5.49 & 7.4 & Loc\\ 
13 & 2200 & 8 & 1.00 & 96 & 80 & 1200 & 121.18 & 0.028 & 1.85 & 3.17 & 6.73 & 2.9 & RaD\\ 
14 & 2200 & 9 & 0.50 & 96 & 80 & 1200 & 101.89 & 0.023 & 1.56 & 2.67 & 5.66 & 10.7 & Loc\\ 
15 & 2200 & 11 & 0.50 & 96 & 96 & 1800 & 107.44 & 0.025 & 1.64 & 2.34 & 3.98 & 36.7 & Loc\\ 
16 & 2200 & 11 & 1.00 & 96 & 96 & 1800 & 134.95 & 0.031 & 2.06 & 2.94 & 5.00 & 35.0 & Loc\\ 
17 & 2400 & 8 & 0.50 & 96 & 96 & 2400 & 103.90 & 0.024 & 1.59 & 2.27 & 2.89 & 5.9 & Loc\\ 
18 & 2400 & 8 & 1.00 & 96 & 96 & 2400 & 127.39 & 0.029 & 1.94 & 2.78 & 3.54 & 3.8 & RaD\\ 
19 & 2500 & 8 & 1.00 & 96 & 80 & 1200 & 129.36 & 0.030 & 1.97 & 3.39 & 7.19 & 1.7 & RaD\\ 
20 & 2500 & 9 & 1.00 & 96 & 80 & 1200 & 137.43 & 0.031 & 2.10 & 3.60 & 7.63 & 10.8 & Int\\ 
21 & 2500 & 11 & 1.00 & 96 & 80 & 1200 & 147.43 & 0.034 & 2.25 & 3.86 & 8.19 & 12.5 & Int\\ 
22 & 2600 & 8 & 1.00 & 96 & 80 & 1200 & 132.19 & 0.030 & 2.02 & 3.46 & 7.34 & 2.8 & StD\\ 
23 & 2600 & 11 & 1.00 & 96 & 80 & 1200 & 148.81 & 0.034 & 2.27 & 3.90 & 8.27 & 11.6 & Int\\ 
24 & 2700 & 8 & 1.00 & 128 & 96 & 1536 & 134.57 & 0.017 & 1.55 & 2.94 & 5.84 & 1.8 & RaD\\ 
25 & 2700 & 9 & 1.00 & 128 & 96 & 1536 & 144.27 & 0.019 & 1.66 & 3.15 & 6.26 & 8.7 & Int\\ 
26 & 2700 & 11 & 1.00 & 128 & 96 & 1536 & 154.13 & 0.020 & 1.77 & 3.36 & 6.69 & 9.2 & Int\\ 
27 & 2800 & 8 & 1.00 & 128 & 96 & 1536 & 136.94 & 0.018 & 1.57 & 2.99 & 5.94 & 2.6 & RaD\\ 
28 & 2800 & 11 & 1.00 & 128 & 96 & 1536 & 159.07 & 0.021 & 1.83 & 3.47 & 6.90 & 9.2 & Int\\ 
29 & 2900 & 8 & 1.00 & 128 & 96 & 1536 & 143.55 & 0.019 & 1.65 & 3.13 & 6.23 & 6.5 & Loc\\ 
30 & 2900 & 9 & 1.00 & 128 & 96 & 1536 & 150.14 & 0.019 & 1.73 & 3.28 & 6.52 & 8.3 & Int\\ 
31 & 2900 & 11 & 1.00 & 128 & 96 & 1536 & 161.83 & 0.021 & 1.86 & 3.53 & 7.02 & 8.5 & Int\\ 
32 & 3000 & 6 & 0.50 & 128 & 128 & 1536 & 114.48 & 0.015 & 1.32 & 1.87 & 4.97 & 3.5 & Loc\\ 
33 & 3000 & 6 & 1.00 & 128 & 96 & 1536 & 129.97 & 0.017 & 1.49 & 2.84 & 5.64 & 1.5 & RaD\\ 
34 & 3000 & 8 & 0.50 & 128 & 96 & 1536 & 118.80 & 0.015 & 1.37 & 2.59 & 5.16 & 4.0 & Int\\ 
35 & 3000 & 8 & 1.00 & 128 & 96 & 1536 & 143.27 & 0.019 & 1.65 & 3.13 & 6.22 & 4.9 & StD\\ 
36 & 3000 & 11 & 0.50 & 96 & 96 & 2800 & 136.95 & 0.031 & 2.09 & 2.99 & 3.26 & 4.6 & Int\\ 
37 & 3000 & 11 & 1.00 & 128 & 96 & 1536 & 166.88 & 0.022 & 1.92 & 3.64 & 7.24 & 4.4 & Int
\end{tabular}
\caption{Simulations in a $L=100D$ long pipe of pulsatile pipe flow performed in this study. In columns find the identification number of the simulation (Case); the flow parameters (\Reynolds, \Womersley, \Amplitude); the radial points $N_{r}$ and half the number of azimuthal and axial Fourier modes $N_{\theta}$ and $N_{x}$ (the total number of physical points is $N_{r} \times 3N_{\theta} \times 3N_{x}$); the maximum $\Reynolds_{\tau}$ and grid discretization in $+$ units; the total number of periods run in the simulation $NT$ and the behavior of the simulation according to the description in section \S \ref{sec:DNSresults}: Rapid decay (RaD), Localized structures (Loc), Stochastic decay (StD) and Highly intermittent state (Int).}
\label{tab:dns}
\end{table} 

\newpage

\begin{table}[h!] 
\begin{tabular}{c|ccc|ccc|ccccc|cc} 
Case & \Reynolds & \Womersley & \Amplitude & $N_{r}$ & $N_{\theta}$ & $N_{x}$ & $\Reynolds_{\tau}$ & $\Delta r_{\min}^{+}$ & $\Delta r_{\max}^{+}$ & $\Delta R \theta^{+}$ & $\Delta x^{+}$ & $NT$ & Behavior \\ \hline 
1 & 2100 & 9 & 0.50 & 80 & 64 & 1152 & 99.35 & 0.032 & 1.81 & 3.25 & 5.75 & 5.8 & Loc\\ 
2 & 2100 & 9 & 0.70 & 80 & 64 & 1152 & 109.47 & 0.036 & 2.00 & 3.58 & 6.33 & 5.8 & Loc\\ 
3 & 2100 & 9 & 0.90 & 80 & 64 & 1152 & 118.58 & 0.039 & 2.17 & 3.88 & 6.86 & 5.8 & StD\\ 
4 & 2100 & 11 & 0.50 & 80 & 64 & 1152 & 104.55 & 0.034 & 1.91 & 3.42 & 6.05 & 6.0 & Loc\\ 
5 & 2100 & 11 & 0.70 & 80 & 64 & 1152 & 116.35 & 0.038 & 2.12 & 3.81 & 6.73 & 7.8 & Loc\\ 
6 & 2100 & 11 & 0.90 & 80 & 64 & 1152 & 126.71 & 0.041 & 2.31 & 4.15 & 7.33 & 6.0 & Loc\\ 
7 & 2100 & 13 & 0.50 & 76 & 64 & 1152 & 109.68 & 0.039 & 2.10 & 3.59 & 6.35 & 9.6 & Loc\\ 
8 & 2100 & 13 & 0.70 & 76 & 64 & 1152 & 122.38 & 0.044 & 2.35 & 4.01 & 7.08 & 9.6 & Loc\\ 
9 & 2100 & 13 & 0.90 & 76 & 64 & 1152 & 133.99 & 0.048 & 2.57 & 4.38 & 7.75 & 9.6 & Loc\\ 
10 & 2100 & 15 & 0.50 & 76 & 64 & 1152 & 114.47 & 0.041 & 2.19 & 3.75 & 6.62 & 13.6 & Loc\\ 
11 & 2100 & 15 & 0.70 & 76 & 64 & 1152 & 128.50 & 0.046 & 2.46 & 4.21 & 7.44 & 13.6 & Loc\\ 
12 & 2100 & 15 & 0.90 & 76 & 64 & 1152 & 140.92 & 0.051 & 2.70 & 4.61 & 8.16 & 13.6 & Loc\\ 
13 & 2100 & 17 & 0.50 & 80 & 64 & 1152 & 119.02 & 0.039 & 2.17 & 3.89 & 6.89 & 11.0 & Loc\\ 
14 & 2100 & 17 & 0.70 & 80 & 64 & 1152 & 134.08 & 0.044 & 2.45 & 4.39 & 7.76 & 11.0 & Loc\\ 
15 & 2100 & 17 & 0.90 & 80 & 64 & 1152 & 147.39 & 0.048 & 2.69 & 4.82 & 8.53 & 11.0 & Loc\\ 
16 & 2300 & 9 & 0.50 & 80 & 64 & 1152 & 104.52 & 0.034 & 1.91 & 3.42 & 6.05 & 4.8 & Loc\\ 
17 & 2300 & 9 & 0.70 & 76 & 64 & 1152 & 114.97 & 0.041 & 2.20 & 3.76 & 6.65 & 4.8 & Loc\\ 
18 & 2300 & 9 & 0.90 & 76 & 64 & 1152 & 124.75 & 0.045 & 2.39 & 4.08 & 7.22 & 4.8 & StD\\ 
19 & 2300 & 11 & 0.50 & 80 & 64 & 1152 & 109.92 & 0.036 & 2.01 & 3.60 & 6.36 & 19.7 & Loc\\ 
20 & 2300 & 11 & 0.70 & 80 & 64 & 1152 & 122.10 & 0.040 & 2.23 & 4.00 & 7.07 & 6.7 & Loc\\ 
21 & 2300 & 11 & 0.90 & 80 & 64 & 1152 & 140.15 & 0.046 & 2.56 & 4.59 & 8.11 & 7.1 & Loc\\ 
22 & 2300 & 13 & 0.50 & 80 & 64 & 1152 & 115.25 & 0.038 & 2.10 & 3.77 & 6.67 & 9.9 & Loc\\ 
23 & 2300 & 13 & 0.70 & 80 & 64 & 1152 & 128.47 & 0.042 & 2.35 & 4.20 & 7.43 & 9.9 & Loc\\ 
24 & 2300 & 13 & 0.90 & 80 & 64 & 1152 & 140.58 & 0.046 & 2.57 & 4.60 & 8.14 & 9.9 & Loc\\ 
25 & 2300 & 15 & 0.50 & 80 & 64 & 1152 & 120.55 & 0.039 & 2.20 & 3.94 & 6.98 & 13.2 & Int\\ 
26 & 2300 & 15 & 0.70 & 76 & 64 & 1152 & 134.89 & 0.048 & 2.59 & 4.41 & 7.81 & 13.2 & Loc\\ 
27 & 2300 & 15 & 0.90 & 76 & 64 & 1152 & 148.14 & 0.053 & 2.84 & 4.85 & 8.57 & 14.8 & Loc\\ 
28 & 2300 & 17 & 0.50 & 80 & 64 & 1152 & 125.86 & 0.041 & 2.30 & 4.12 & 7.28 & 20.0 & Loc\\ 
29 & 2300 & 17 & 0.70 & 80 & 64 & 1152 & 141.07 & 0.046 & 2.58 & 4.62 & 8.16 & 20.0 & Loc\\ 
30 & 2300 & 17 & 0.90 & 80 & 64 & 1152 & 154.80 & 0.050 & 2.83 & 5.07 & 8.96 & 16.0 & Loc\\ 
31 & 2500 & 9 & 0.50 & 80 & 64 & 1152 & 109.47 & 0.036 & 2.00 & 3.58 & 6.34 & 4.9 & Loc\\ 
32 & 2500 & 11 & 0.50 & 80 & 64 & 1152 & 115.18 & 0.038 & 2.10 & 3.77 & 6.67 & 5.8 & Loc\\ 
33 & 2500 & 17 & 0.50 & 80 & 64 & 1152 & 133.98 & 0.044 & 2.45 & 4.38 & 7.75 & 17.5 & Int\\ 
34 & 2500 & 17 & 0.70 & 80 & 64 & 1152 & 149.59 & 0.049 & 2.73 & 4.90 & 8.66 & 17.5 & Int
\end{tabular} 
\caption{Simulations of pulsatile pipe flow using the master-slave method described in section \S \ref{sec:master-slaveresults} performed in this study. All the simulations correspond to a $L=100D$ long pipe. In columns find the identification number of the simulation (Case); the flow parameters (\Reynolds, \Womersley, \Amplitude); the radial points $N_{r}$ and half the number of azimuthal and axial Fourier modes $N_{\theta}$ and $N_{x}$ (the total number of physical points is $N_{r} \times 3N_{\theta} \times 3N_{x}$); the maximum $\Reynolds_{\tau}$ and grid discretization in $+$ units; the total number of periods run in the simulation $NT$ and the behavior of the master simulation according to the description in section \S \ref{sec:DNSresults}: Rapid decay (RaD), Localized structures (Loc), Stochastic decay (StD) and Highly intermittent state (Int).}
\label{tab:MSdns}
\end{table} 

\newpage

\appendix

\section{Detailed description of master-slave DNS}\label{ap:master_slave}
In this appendix we provide a detail derivation of the slave profile and the methods we use to integrate our master-slave DNS. Note that, in this appendix, we normalize $r$ using the radius $R$, and not the pipe diameter.
The slave profile is defined so it complies with a series of conditions.
\subsection{Condition 1: boundary condition}
The slave profile must comply with the no-slip boundary condition and thus vanish at the wall:
\begin{equation}
U_{S} \left(r=R,t\right)=0\text{.}
\label{eq:CH8_cond1}
\end{equation}
\subsection{Condition 2: time dependency and maximum energy}
The slave profile must be time dependent. Its bulk velocity $\bar{U}_{S}\left(t\right)$:
\begin{equation}
\bar{U}_{S}\left(t\right)=\frac{2}{R^{2}}\int_{0}^{R}  U_{S} r \mathrm{d}r \text{,}
\end{equation}
is set equal to
\begin{equation}
\bar{U}_{S}\left(t\right)= \sqrt{\frac{3 E_{L}\left(t\right)}{2}}\text{,}
\label{eq:CH8_cond2}
\end{equation}
being:
\begin{equation}
E_{L} \left(t \right)=\frac{1}{\pi R^{2}}\int_{0}^{2\pi}\int_{0}^{R} \frac{1}{2} U_{SW}^{2} r\mathrm{d}r\mathrm{d}\theta \text{,}
\end{equation}
the kinetic energy of the laminar pulsatile pipe flow $U_{SW}\left(r,t\right)$. With this condition we ensure that the energy of the profile is always equal or smaller than the corresponding laminar $U_{SW}$ profile.
\subsection{Condition 3: monotonic shear}
The average shear:
\begin{equation}
S =  \frac{2}{R^{2}}\int_{0}^{R} \frac{1}{2} \left( \frac{\partial U_{S}}{\partial r}\right) ^{2} r\mathrm{d}r \text{,}
\label{eq:CH8_cond3}
\end{equation}
of the profile must be minimum. Given conditions \eqref{eq:CH8_cond1} and \eqref{eq:CH8_cond2}, by minimizing $S$, we obtain profiles whose shear monotonically decreases from the wall to the center-line of the pipe, without inflection points. 
\subsection{Laminar slave mean profile}
The parabolic profile:
\begin{equation}
U_{S_{0}}\left(r,t\right)=2 \bar{U}_{S}\left(t\right) \left(1- \left(\frac{r}{R}\right)^{2}\right) \text{,}
\label{eq:CH8_lam_slave}
\end{equation}
whose energy is exactly $E_{L}\left(t\right)$, i.e. complies with these three initial requirements. By dropping the time dependence of $\bar{U}_{S}$ and $E_{L}$ in the notation and setting $R=1$, one finds:
\begin{equation}
\int_{0}^{1} U_{S_{0}}^{2} r \mathrm{d}r=4\bar{U}_{S}^{2}\int_{0}^{1}\left(1-r^{2}\right)^{2}r \mathrm{d}r=4\bar{U}_{S}^{2}\left[\frac{r^{2}}{2}-\frac{2 r^{4}}{4}+ \frac{r^{6}}{6} \right]_{0}^{1}=\frac{4 \bar{U}_{S}^{2}}{6}=E_{L} \text{.}
\end{equation}
In the case of turbulent flow, the energy of the mean profile is smaller than the laminar one, \citep{kuhnen}. To enforce this condition, and avoid introducing excess kinetic energy in the slave simulations we use an additional constraint.
\subsection{Condition 4: energy of the slave mean profile}
The slave mean profile must have the same energy as the master mean profile:
\begin{equation}
\frac{2}{R^{2}}\int_{0}^{R} \frac{1}{2} U_{S} ^{2} r\mathrm{d}r = E_{M} \left( t \right)= \frac{2}{R^{2}}\int_{0}^{R} \frac{1}{2} \lla u_{x} \rra_{\theta,x} ^{2} r\mathrm{d}r \text{.}
\label{eq:CH8_cond4}
\end{equation}
If $E_{M} \equiv E_{L}$, the flow in the master simulation is laminar, $U_{S}\equiv U_{S_{0}}$, and the energy of the slave mean profile will be maximum. Otherwise $E_{M}<E_{L}$, and the resultant $U_{S}$ is blunted, as seen in figure~\ref{fig:fig1}.
\subsection{Method of small variations}
We can express mathematically conditions \eqref{eq:CH8_cond2}--\eqref{eq:CH8_cond4} in a functional:
\begin{equation}
\mathcal{S}=2\int_{0}^{R} \mathcal{L} \left(r,U_{S},U_{S}'\right) \mathrm{d} r \text{,}
\end{equation}
to be minimised, being $U_{S}'=\frac{\partial U_{S}}{\partial r}$,
\begin{equation}
\mathcal{L}= \frac{1}{2} {U'}_{S}^{2} r + \lambda_{L} \left(U_{S} r - \frac{\bar{U}_{S}}{2R} \right) + \mu_{L} \left(\frac{1}{2}U_{S}^{2} r - \frac{E_{M}}{2 R} \right)
\label{eq:CH8_Lagrangian}
\end{equation}
the Lagrangian, and $\lambda_{L}$ and $\mu_{L}$ two Lagrange multipliers. We have dropped the time dependence of $\bar{U}_{S}$, $E_{M}$, $U_{S}$ and therefore $\mathcal{L}$, $\lambda_{L}$ and $\mu_{L}$ in the notation for clarity.

From the method of small variations, one can find the function $U_{S}$ that minimizes $\mathcal{L}$ by solving the Euler-Lagrange equation:
\begin{equation}
\frac{\partial \mathcal{L}}{\partial U_{S}}- \frac{\partial}{\partial r} \frac{\partial \mathcal{L}}{\partial  U_{S}'}=0 \text{.}
\end{equation}
In this case, one finds:
\begin{equation}
\lambda_{L} r + \mu_{L} U_{S} r  - \frac{\partial}{\partial r}\left(r U_{S}' \right)=0 \text{,}
\end{equation}
that, after rearranging, results in the partial differential equation:
\begin{equation}
\frac{\partial ^{2} U_{S}}{\partial r^{2}} + \frac{1}{r} \frac{\partial U_{S}}{\partial r} - \mu_{L} U_{S} = \lambda_{L} \text{,}
\label{eq:CH8_Eu_Lag}
\end{equation}
with the boundary condition~\eqref{eq:CH8_cond1}:
\begin{equation}
U_{S}\left(r=R\right) = 0 \text{.}
\end{equation}
The homogeneous part of equation~\eqref{eq:CH8_Eu_Lag} can be turned into a modified Bessel's equation of order 0:
\begin{equation}
r^{2} \frac{\partial ^{2} U_{S}}{\partial r^{2}} + r \frac{\partial U_{S}}{\partial r} - \mu_{L} U_{S}r^{2} = 0 \text{,}
\label{eq:CH8_hom}
\end{equation}
being the modified Bessel's equation of order $\nu_{B}$:
\begin{equation}
x^{2} y'' + x y' - \left( a^{2} x^{2} + \nu_{B}^{2} \right) y=0 \text{,}
\end{equation}
with solution
\begin{equation}
y\left(x\right)= A I_{\nu_{B}} \left(a x\right) + B K_{\nu_{B}} \left(a x\right) \text{.}
\label{eq:CH8_bessel}
\end{equation}
Here $I_{\nu_{B}}$ and $K_{\nu_{B}}$ are the modified Bessel functions of order $\nu_{B}$, of the first and second kind, and $A$ and $B$ integration constants. After comparing equation \eqref{eq:CH8_hom} with \eqref{eq:CH8_bessel}, \mod{$a=\sqrt{\mu_{L}}$} and $\nu_{B}=0$. The solution is written as:
\begin{equation}
U_{S}= A I_{0}\left( \sqrt{\mu_{L}} r \right) + B K_{0} \left(\sqrt{\mu_{L}} r\right) \text{.}
\end{equation}
Since $K_{0}$ diverges at $r=0$, $B=0$. Regarding the particular solution, the constant $U_{S}=C$:
\begin{equation}
- \mu_{L} C= \lambda_{L} \rightarrow C=- \frac{\lambda_{L}}{\mu_{L}} \text{,}
\end{equation}
is tried, yielding 
\begin{equation}
U_{S}=  A I_{0}\left( \sqrt{\mu_{L}} r \right) - \frac{\lambda_{L}}{\mu_{L}} \text{.}
\end{equation}
One can determine the constant $A$ from the boundary condition, at $r=R\equiv 1$, and find:
\begin{equation}
U_{S}=\frac{\lambda_{L}}{\mu_{L}}\left( \frac{I_{0}\left( \sqrt{\mu_{L}} r \right)}{I_{0}\left( \sqrt{\mu_{L}} \right)} -1 \right) \text{.}
\label{eq:CH8_slave_prof}
\end{equation}

\modd{At each time step a Newton--Raphson method is used to find the correct $\mu_{L}$ and $\lambda_{L}$, that allow $U_{S}$ to fulfill conditions~\eqref{eq:CH8_cond3} and \eqref{eq:CH8_cond4}. }

\subsection{Computational set-up of master-slave DNS}
We initialize each pair of master-slave DNS with the same flow field. This flow field is the resultant field of a previous DNS at similar flow parameters, and has only one localized turbulent structure. In order to integrate simultaneously the master and slave simulations, our code performs the following sub-steps at each time step:
\begin{enumerate}
\item The master simulation is integrated one time step.
\item Using the instantaneous mean profile $U_{M} = \lla u_{x} \rra_{\theta,x}$ of the master, it computes the instantaneous energy $E_{M}$, see eq.~\eqref{eq:CH8_cond4}. In our pseudo-spectral code, $U_{M}=\lla u_{x} \rra_{\theta,x}$, corresponds to the $\left(0,0\right)$ Fourier mode of the axial velocity.
\item Using the corresponding laminar pulsatile pipe flow kinetic energy $E_{L}$ it computes the desired $\bar{U}_{S}$, see eq.~\eqref{eq:CH8_cond2}.
\item It uses a Newton-Raphson method to compute the $U_{S}$ profile, eq.~\eqref{eq:CH8_slave_prof}, that complies with the desired $\bar{U}_{S}$ and $E_{M}$. 
\item It overwrites the mean profile of the slave simulation and imposes $U_{S}$ instead. (In the code, it overwrites the $\left(0,0\right)$ Fourier mode of the axial velocity of the slave simulation.)
\item It integrates one time step the slave simulation, ignoring the evolution of its mean profile.
\end{enumerate}

\subsubsection{The pipe length}
The artificial profiles $U_{S}$ depend on the selected length of the pipe $L$. For an infinitely long pipe $L\rightarrow \infty$, with a single localized turbulent puff, the mean profile of the master simulation will tend to the laminar profile $\lla u_{x} \rra_{\theta,x} \rightarrow U_{SW}$. This means that the energy of the mean profile $E_{M} \rightarrow E_{L}$ and therefore $U_{S} \rightarrow U_{S_{0}}$. The Newton-Raphson method works better as long as $E_{M} \approx E_{L}$. However the computational cost increases as the length of the pipe increases. A good compromise is found, by setting a length of $L=100D$. 

\section{ Detailed description of the EBM}\label{ap:ebm}
In this appendix we first describe the BM \citep{barkley2015rise}. We then justify the changes we use to extend the BM to the EBM and we explain the methods by which we numerically integrate the EBM. At the end of this appendix we also comment on the limitations of the EBM.

\subsection{The original BM}
The original BM considers two one-dimensional time-dependent variables $q\left(x,t\right)$ and $u\left(x,t\right)$. The former corresponds to the turbulence intensity at each axial location and time. According to \citet{barkley2015rise} $q$ represents some form of the cross-section integral of the cross-section kinetic energy, so $q \geq 0$. The variable $u$ is a proxy of the state of the mean shear at each axial location and time, represented by the axial center-line velocity in the model. The center-line velocity $u$ is bounded between the bulk velocity $\bar{U}=1$ and the laminar center-line velocity $U_{c}=2$ so $1 \leq u \leq 2$. The evolution equations of $q$ and $u$ in the BM read \citep{barkley2015rise}:
\begin{equation}
    \frac{\partial q}{\partial t} =-\left(u-\zeta \right)\frac{\partial q}{\partial x} + f\left(q,u \right) + D_{q}\frac{\partial^{2} q}{\partial x^{2}} + \sigma \tau \left(t,x\right) q \text{,} \label{eq:SPDE_q} 
\end{equation}
\begin{equation} 
    \frac{\partial u}{\partial t}=-u\frac{\partial u}{\partial x} + g\left(q,u \right) \text{,}  \label{eq:PDE_u} 
\end{equation}
with
\begin{equation}
    f\left(q,u\right) =q \left[r+u-U_{c}-\left(r+\delta \right) \left(q -1 \right)^{2} \right] \text{,} \label{eq:loc_q} 
\end{equation}
\begin{equation}
    g\left(q,u\right) = \epsilon \left(U_{c}-u \right) + 2 \epsilon \left(\bar{U}-u \right)q \text{.} \label{eq:loc_u}
\end{equation}
$\tau \left(t,x\right)$ is white Gaussian noise in space and time, and 
\begin{equation}
r=\frac{\Reynolds - R_{0}}{R_{1} - R_{0}} \text{,}
\label{eq:ch5_rvsRe}
\end{equation}
is a control parameter that represents a rescaled Reynolds number. The parameter $R_{0}$ corresponds to the first $\Reynolds$ at which puffs survive for long times $R_{0}=1920$. The parameter $R_{1}$ corresponds to the first $\Reynolds$ at which puffs elongate into slugs $R_{1}=2250$.

Overall, the model has seven parameters: $\zeta$, $D_{q}$, $\sigma$, $\delta$, $R_{0}$, $R_{1}$, $\epsilon$. When correctly fitted, see table~\ref{tab:params}, the model perfectly reproduces the front speed of puffs and slugs in SSPF \citep{barkley2015rise,chen2022}. For a comprehensive description of the model and its underlying ideas, the reader is referred to \citep{barkley2016} and references therein. 

\begin{table}
\centering
\begin{tabular}{cccccccc}
              &$R_{0}$ & $R_{1}$ & $\zeta$ & $D_{q}$  & $\sigma$ & $\delta$ & $\epsilon$ \\ \hline
\textbf{BM} & 1920    & 2250     & 0.79    & 0.13 & $\leq 0.5$& 0.1      & 0.2         \\ 
\textbf{EBM}& 1920    & 2250     & 0.79    & 0.13 & $0.2\leq \sigma \leq0.85$ & 0.1 & 0.1
\end{tabular}
\caption{BM parameters as described in \citet{barkley2015rise} and the value of parameters used in the EBM.}
\label{tab:params}
\end{table}

\subsubsection{Time scale of the model}
In order to compare the model and DNS/experiment front speeds, \citet{barkley2015rise} proposes the velocity scale difference:
\begin{equation}
\psi=2 \left(C_{0} -C_{1} \right)=0.28 \text{.}
\label{eq:timescale}
\end{equation}
Here $C_{0}$ is the front speed velocity of puffs at $\Reynolds=R_{0}$ and $C_{1}$ the front speed of puffs at $\Reynolds=R_{1}$.

This velocity scale difference can also be understood as a time scale difference between model and DNS/experiments. In particular, an advective time unit $\left(D/U\right)$ in DNS/experiments, corresponds to $\psi=0.28$ time units in the model. This scale difference is important for the pulsatile case, where a pulsation period of length $T=\frac{\pi \Reynolds}{2 \Womersley^{2}} $ in advective time units in DNS/experiments, corresponds to a period of length $T^{*}=0.28 T$ in the model.

\subsection{Equations of the EBM}
In order to adapt the BM to pulsatile pipe flow we introduce several changes. The equations of the EBM read:
\begin{equation}
    \frac{\partial q}{\partial t} =-\left(u-\zeta \bar{U}\left(t\right) \right)\frac{\partial q}{\partial x} +f_{EBM}\left(q,u \right) + D_{q}\frac{\partial^{2} q}{\partial x^{2}} + \sigma \left(\Reynolds \right) \tau \left(t,x\right) q \text{,}
\label{eq:SPDE_q_EBM}
\end{equation}
\begin{equation}
    \frac{\partial u}{\partial t} =-u\frac{\partial u}{\partial x} + g_{EBM}\left(q,u\right) \text{,} \label{eq:PDE_u_EBM} 
\end{equation}
with
\begin{equation}
    f_{EBM}\left(q,u\right) =q \left[r\bar{U}\left(t+\phi\right)+\gamma \lambda \left( t \right) + u-U_{c}\left(t\right)-\left(r\bar{U}\left(t+\phi\right)+\delta \right) \left(q -1 \right)^{2} \right]\text{,}
\label{eq:loc_q_EBM}
\end{equation}
\begin{equation}
    g_{EBM}\left(q,u\right) = \epsilon \left(U_{c}\left(t\right)-u \right) + 2 \epsilon \left(\bar{U}\left(t\right)-u \right)q + P_{G}\left(t\right) + F_{v_{0}}\left(t\right)\text{,} \label{eq:loc_u_EBM}
\end{equation}
where $\bar{U}$, $U_{c}$, $P_{G}$ and $F_{v_{0}}$ are the corresponding bulk velocity, the laminar center-line velocity, the pressure gradient and the viscous force at the center-line of the pipe. Note that, different to the BM, these quantities are time dependant. Moreover, $\phi$ corresponds to a phase lag between the turbulence intensity and the bulk velocity. The product of $\gamma$ and $\lambda$ models the effect of inflection points in the $U_{SW}$ profile on $q$. In order to better fit the model results to the DNS results we set $\epsilon=0.1$ and change $\sigma$ depending on \Reynolds:
\begin{equation}
\sigma= \frac{6}{5}\cdot \frac{\left(\Reynolds-1933 \right )}{1000} \text{,}
\label{eq:sigma_EBM}
\end{equation}
with a lower limit of $\sigma \geq 0.2$, so there are always some stochastic behaviors, and an upper limit of $\sigma \leq 0.85$, so the stochastic term is never too dominant. In table~\ref{tab:params}, we list the parameter values we use for the EBM throughout the paper. In what follows we justify the changes introduced in the EBM.

\subsection{Extensions to the $u$ equations}
In pulsatile pipe flow, the laminar center-line velocity $U_{c} \left(t \right)$ and bulk velocity $\bar{U} \left( t \right)$ are functions of time. While the bulk velocity is set by the pulsation, the evolution of $U_{c}\left( t \right)$ can be obtained from the NSE. First we assume laminar flow, $\pmb{u}\left(r,\theta,x,t\right)\rightarrow\left(0,0,U_{SW}\left(r,t\right) \right)$ and, from the NSE, obtain an equation for the evolution of the (SW) laminar profile:
\begin{equation}
\frac{\partial U_{SW}}{\partial t}= P_{G}\left(t\right) + F_{visc}\left(r\right) \text{,}
\end{equation}
Here $P_{G}\left(t\right)$ is the pressure gradient that drives the flow at the desired bulk velocity $\bar{U} \left( t \right)$, and:
\begin{equation}
F_{visc}\left(r\right)= \frac{1}{\Reynolds} \left( \frac{\partial^{2}U_{SW}}{\partial r^{2}} + \frac{1}{r} \frac{\partial U_{SW}}{\partial r} \right) \text{,}
\end{equation}
the viscous forces. At the center-line of the pipe $r \rightarrow 0$,
\begin{equation}
F_{v_{0}} \left(t \right)=\lim_{r \to 0} F_{visc} = \lim_{r \to 0} \left[ \frac{1}{\Reynolds} \left( \frac{\partial^{2}U_{SW}}{\partial r^{2}} + \frac{1}{r} \frac{\partial U_{SW}}{\partial r} \right) \right] \text{.}
\label{eq:limFvisc}
\end{equation}
If one applies L'Hopital's rule to the limit in equation~\eqref{eq:limFvisc},
\begin{equation}
F_{v_{0}} \left(t \right) =  \frac{2}{\Reynolds} \frac{\partial^{2}U_{SW}}{\partial r^{2}} \text{,}
\end{equation}
and, being $U_{c}\left(t\right)=U_{SW}\left(t,r=0\right)$,
\begin{equation}
\frac{\partial U_{c}}{\partial t}=P_{G}\left(t\right) + F_{v_{0}}\left(t\right) \text{.}
\label{eq:center-line}
\end{equation}
Note that the time average of the right hand side of equation~\eqref{eq:center-line} yields $\left \langle P_{G}\left(t\right) + F_{v_{0}}\left(t\right) \right \rangle_{t}=0$. Therefore in the BM there is no need to consider these terms as they cancel each other at each time step. In the case of pulsatile pipe flow however, the equilibrium of forces described in equation~\eqref{eq:center-line} must be included in the model, see eq.~\eqref{eq:loc_u_EBM}. 

\subsection{Extensions to the $q$ equations}
In the original BM, \citet{barkley2015rise} assumed that the turbulence intensity is advected at the center-line velocity $u$, corrected with the parameter $\zeta$. In the case of the EBM we multiply the parameter $\zeta$ in equation~\eqref{eq:SPDE_q_EBM} by the bulk velocity $\bar{U}$ to account for the effect of the pulsation. In the rest of this subsection we justify other changes to the evolution equations of $q$ in the EBM.

\subsubsection{Phase lag $\phi$}
In previous studies of pulsatile pipe flow, a phase lag between the driving bulk velocity and turbulence has been reported \citep{weng2016}. In our DNS we observe a time delay between the maximum integrated turbulence intensity $\left \langle \omega_{x}^{2} \right \rangle_{r,\theta,x}$ and the bulk velocity $\bar{U}$, see fig.~\ref{fig:fig3}, that is mainly dependent on \Womersley. We find that, for the model purposes, the analytical phase lag $\phi \left(\Womersley\right) \approx 32.34^{\circ}+35.17^{\circ} \arctan  \left( 0.75 \left( \Womersley -2 \right) \right)$ between the pressure gradient and laminar profile, first derived by Womersley \cite{womersley1955method}, is a good approximation to this phase difference.

\subsubsection{The effect of the inflection points: $\lambda$ and $\gamma$}
At $5 \lesssim \Womersley \lesssim 19$ and $\Amplitude\gtrsim 0.5$, the laminar profile of pulsatile pipe flow is very different to the parabolic profile of the steady case, and is instantaneously unstable at certain phases of the period \citep{moron2022}. The center-line velocity $u$ alone is not able to capture these features of the mean shear. In the EBM the instantaneous shape of the mean shear and its effects are modeled by adding to equation~\eqref{eq:loc_q_EBM} the term $+ \gamma \lambda \left(t\right)$. 

\begin{figure}
\centering
\includegraphics[width=\linewidth, trim=0mm 0mm 0mm 0mm, clip=true]{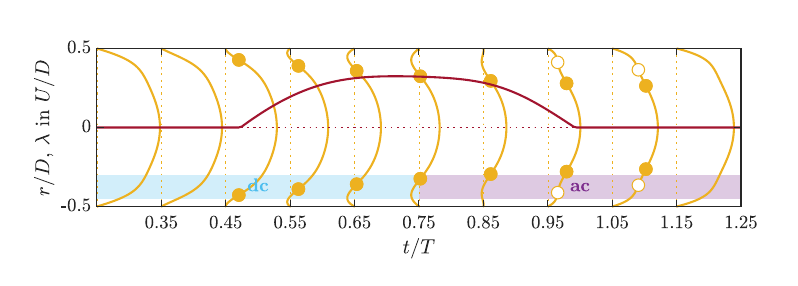}
\caption{Laminar profile and instantaneous maximum eigenvalue $\lambda_{\max}$ according the instantaneous stability analysis by \citet{moron2022}. In yellow the instantaneous laminar profiles $U_{SW}$ at $\Reynolds=2100$, $\Womersley=11$ and $A=0.9$. To not interfere with one another the profiles are scaled using a scalar with arbitrary units so the all time maximum is smaller than $t/T=\num{0.15}$, since only the development of $U_{SW}$ in time is of interest. With points find the existence and position $r_{i}$ of inflection points in the profile. Filled points correspond to inflection points that also satisfy the Fj{\o}rtoft criterion locally $\frac{\partial^2 U_{SW}}{\partial r^2}\left(U_{SW}-U_{SW}\left(r_{i} \right ) \right )<0$. In red, the parameter $\lambda=\max\left(0,\lambda_{\max}\right)$ used in the EBM.} 
\label{fig:ap}
\end{figure}

Here $\lambda \left(t\right)$ represents how instantaneously linearly unstable the laminar profile is, and is always $\lambda\geq 0$. We compute it using the method described in \citet{moron2022}. We compute the eigenvalues of the laminar profile at certain time steps as if the profile was instantaneously steady, using the method proposed by \citet{Meseguer}. We then assume that the maximum eigenvalue is continuous in time and construct $\lambda \left(t \right)$. We then force $\lambda$ to be $\lambda\geq 0$. For more details the reader is referred to \citet{moron2022}. For an example of $\lambda$ in the model, see figure~\ref{fig:ap}. Note that, during the phases of the period where there are inflection points in the profile, $\lambda$ is most of the time $\lambda>0$. 

The parameter $\gamma$ sets the effect $\lambda$ has on the growth of $q$ in equation~\eqref{eq:loc_q_EBM}. It models the quality of the quasi-steady assumption used to compute $\lambda$. It scales with the length of the period in terms of flow units $T=\frac{\pi \Reynolds}{2 \Womersley^{2}}$. The idea is that, the longer the period is, the slower the mean shear evolves with respect to the turbulent structures. We find a good compromise with:
\begin{equation}
\gamma = \min \left( 1,0.28 \log \left(T\right)\right) \text{.}
\label{eq:gamma_EBM}
\end{equation}
We impose an upper boundary of $\gamma=1$ so the dynamics of the model are not dominated by $\gamma \lambda$.

\subsubsection{The noise intensity $\sigma$}
\begin{figure}
\centering
\includegraphics[width=\textwidth, trim=0mm 0mm 0mm 0mm, clip=true]{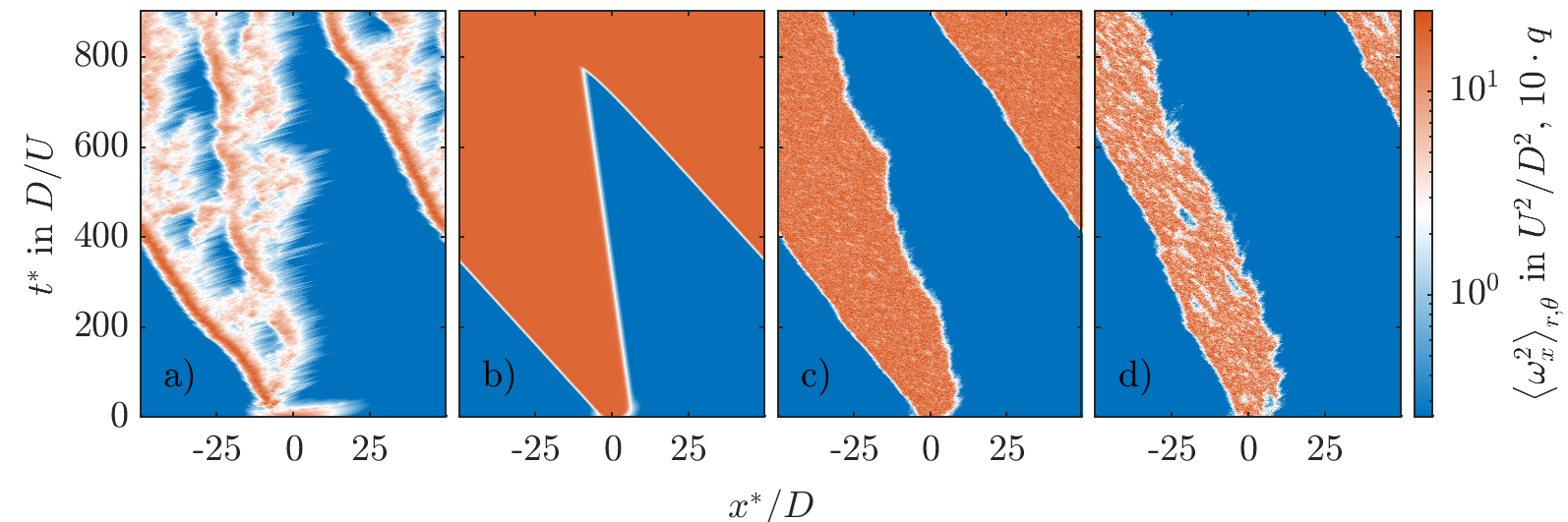}
\caption{Space-time diagrams of the cross section integral of axial vorticity squared of DNS (a) and $10 \cdot q$ of the BM (b, c, d). The results correspond to DNS and BM simulations in a $100D$ long pipe, at $\Reynolds=2400$, initialized with a localized perturbation of length $5D$. The figure is presented with respect to a moving frame $x^{*}$, moving with the bulk velocity $\bar{U}$, and time $t^{*}$ in advective time units for DNS, and model units for the BM. Subplot b) corresponds to a simulation of the BM with the parameters listed in table~\ref{tab:params} and $\sigma=0$; c) $\sigma=0.5$ and d) $\sigma=0.7$.}
\label{fig:ap2}
\end{figure}

The original BM does not capture all the chaotic behaviors of localized turbulence in the \Reynolds regime at $2250 \lesssim \Reynolds \lesssim 2500$. In this regime, according to the BM, puffs elongate into slugs filling the whole pipe with turbulence, as seen in figure~\ref{fig:ap2}b, c and d. However, in full DNS, at these \Reynolds the flow usually reaches a highly heterogeneous state, where localized turbulent patches coexist with laminar flow patches, see fig.~\ref{fig:ap2}a. This behavior can only be approximated by radically increasing the noise parameter $\sigma$ in the model, see fig.~\ref{fig:ap2}d, but at the same time loosing the good agreement between model and DNS/experiments front speeds.

This limitation of the BM turns out to be even more detrimental in the case of the EBM as \Reynolds increases. We find that a noise intensity $\sigma$ that scales linearly with \Reynolds in eq.~\ref{eq:sigma_EBM} produces better results. The parameter has a lower bound so the model has some degree of stochastic behavior at low \Reynolds, and an upper bound so the noise term is never more dominant than the rest of the dynamics of the model. 

\subsubsection{The parameter $\epsilon$}
In the original work of \citet{barkley2016}, they suggest that $\epsilon$ should be inversely proportional to $\Reynolds$. But, since changing this parameter does not have a huge impact on turbulence front speeds and survival in the case of SSPF, they keep it constant. In the case of pulsatile pipe flow, in order to find a better match with the DNS results, it should be slightly decreased. This is justified since, the maximum $\Reynolds_{\max} = \left( 1+A\right) \Reynolds$ is, in the worst case scenario, two times the mean \Reynolds. Therefore $\epsilon$ is set to half its BM value in the EBM, see table~\ref{tab:params}.

\subsection{Computational set-up of the EBM}
Equations \eqref{eq:SPDE_q_EBM} and \eqref{eq:PDE_u_EBM} are integrated following \citet{barkley2015rise}. The second order derivatives are discretized with second order central finite differences, and the first order derivatives with a first order upwind scheme. The system is integrated using an explicit Euler method, with a time step size $\Delta t=0.0025 $. The results here correspond to a pipe of length $L=100$ and a uniform grid spacing $\Delta x=0.5$. The stochastic term is modelled as white Gaussian noise in space and time. All the EBM simulations are initialized with a localized $5D$ long disturbance with magnitude $q\leq0.5$ at the initial time $t_{0}=T^{*}/2$.

In order to prepare all the variables needed to integrate the EBM the following algorithm has been implemented. After selecting the desired $\Amplitude$, $\Reynolds$ and $\Womersley$ the code first numerically integrates the corresponding laminar profile to obtain all the time dependent parameters: $\bar{U}$, $U_{c}$, $P_{G}$, $F_{v_{0}}$ and $\lambda$. It then computes the phase shift angle $\phi \left(\Womersley\right)$, the parameter $\sigma$ using eq.~\eqref{eq:sigma_EBM} and scales the pulsation period to adapt it to the time scale of the model using equation~\eqref{eq:timescale}. Finally it integrates the equations~\eqref{eq:SPDE_q_EBM}, \eqref{eq:PDE_u_EBM}, \eqref{eq:loc_q_EBM} and \eqref{eq:loc_u_EBM}.

\subsection{The EBM at $\Amplitude=0.0$}
Note that if one sets $\Amplitude=0$ in the EBM, except for the new definitions of $\sigma$ and $\epsilon$, one recovers the original BM. At $\Amplitude=0$ the parameters derived from the laminar flow are constant in time: $U_{c}=2$, $\bar{U}=1$, $P_{G}+F_{v_{0}}=0$ and $\lambda=0$ and the EBM equations~\eqref{eq:SPDE_q_EBM}--\eqref{eq:loc_u_EBM} return to  the BM equations~\eqref{eq:SPDE_q}--\eqref{eq:loc_u}.

\subsection{Limitations of the EBM} \label{sec:limitsEBM}
When correctly fitted the EBM captures the dynamics of pulsatile pipe flow in a broad parametric regime, but it has some limitations that need to be mentioned. Some of its limitations are inherited from the original BM. One is the intermittent behavior of turbulence at some flow parameters, that is only captured if $\sigma$ is scaled with \Reynolds. The other is the magnitude of $q$ in the core of the slugs, that the BM tends to overestimate, \citep{barkley2016}. This results in some qualitative differences in the structures observed in the model and the DNS.

New to the EBM is the sensitivity to the parameter $\gamma$. The EBM is expected to work worse as $\Amplitude$ increases. At higher $\Amplitude$, and $\Womersley >5$, the laminar profile is very different to the simple parabolic profile of steady pipe flow, see fig.~\ref{fig:ap}, and the state of the mean shear can no longer be captured with one variable $u$. Here, instead of introducing more variables in the analysis, we find a a work-around by using the instantaneous linear instability $\lambda$, and the parameter $\gamma$. Both work reasonably well, as long as $\gamma$ is correctly fitted. But as soon as $\gamma$ is changed, puffs either decay or elongate when they should not.

Also, due to the definition of $\gamma$ the model overestimates the lifetime of puffs at certain flow parameters. In particular, according to the EBM, puffs at $10 \lesssim \Womersley \lesssim 15$, $\Amplitude= 1$ and $1800 \lesssim \Reynolds \lesssim 2050$ survive the pulsation. This is obviously not observed in DNS of pulsatile pipe flow, where at $\Reynolds \leq 2000$ and $\Amplitude=1$ puffs tend to decay independently of the pulsation frequency, \citep{moron2022}. Moreover, at these flow parameters, the EBM is clearly dominated by the parameter $\lambda$, and therefore by $\gamma$. We do not show it here as we do not consider these flow parameters in our analysis, but we believe it to be fair to at least mention these errors.

The EBM considers that, as long as $\lambda>0$, turbulence can always make use of the instantaneous linear instability to grow. However this may not be the case in a full DNS. At a given time step, the mean profile of a DNS can be highly perturbed. In this case, puffs would not have the chance to take advantage of the linear instability to grow, and could even decay. This feature should be considered in future versions of the model.

\end{document}